\documentclass[english,12pt,aps,prd,a4paper,preprintnumbers,floatfix,nofootinbib,showpacs,superscriptaddress, notitlepage]{revtex4-1} 
 \pdfoutput=1
\usepackage[usenames,dvipsnames]{color}  
\usepackage{graphicx}
\usepackage[utf8]{inputenc}
\usepackage{caption}
\usepackage{subcaption}
\usepackage{amsmath}
\usepackage{amssymb}
\usepackage[colorlinks=true,citecolor=darkred,urlcolor=darkred, pdfborder={0 0 0}]{hyperref}
\usepackage[normalem]{ulem}

\usepackage[T1]{fontenc}
\usepackage{array}
\usepackage{booktabs}
\usepackage{mathrsfs}
\usepackage{multirow}
\usepackage{tabularx}
\usepackage[utf8]{inputenc}
\definecolor{darkred}{rgb}{0.6,0,0}

\definecolor{linkcolor}{rgb}{0,0,0.5}

\def\gsim{\raise0.3ex\hbox{$\;>$\kern-0.75em\raise-1.1ex\hbox{$\sim\;$}}}
\def\lsim{\raise0.3ex\hbox{$\;<$\kern-0.75em\raise-1.1ex\hbox{$\sim\;$}}}

\def\beqn#1{\begin{equation}\label{#1}}
\def\eeqn{\end{equation}}

\def\beqa#1{\begin{eqnarray}\label{#1}}
\def\eeqa{\end{eqnarray}}

\def\Z2{$\mathcal{Z_2}$}

\newcommand {\ignore}[1]{}

\def\cevns{CE$\nu$NS~}
\def\inc{I$\nu$NS~}

\def\321{$\mathrm{SU(3) \otimes SU(2) \otimes U(1)}$ }

\def\be{\begin{equation}}

\def\ee{\end{equation}}
\def\barr{\begin{array}}
\def\earr{\end{array}}

\def\nn8{\nonumber\\[2pt]}
\def\l{\left}
\def\r{\right}
\def\dis{\displaystyle}
\def\ed{\end{document}}

 \newcommand{\AddrIOANNINA}{%
 Division of Theoretical Physics, University of  Ioannina, GR 45110 Ioannina, Greece}

\begin{document}
\sloppy
 
\title{\boldmath \color{BrickRed} 
 Elastic and inelastic scattering of neutrinos and weakly interacting massive particles on nuclei}


\bibliographystyle{unsrt}   

\author{R. Sahu}\email{rankasahu@gmail.com}\affiliation{National Institute
of Science and Technology, Palur Hills, Berhampur 761 008, Odisha, India}
\author{D.K. Papoulias}\email{d.papoulias@uoi.gr}\affiliation{\AddrIOANNINA}
\author{V.K.B. Kota}\email{vkbkota@prl.res.in}\affiliation{Physical 
Research Laboratory, Ahmedabad 380 009, India}
\author{T.S. Kosmas}\email{hkosmas@uoi.gr}\affiliation{\AddrIOANNINA}

\begin{abstract}
The event rates for WIMP-nucleus and neutrino-nucleus scattering processes, 
expected to be detected in ton-scale rare-event detectors, are investigated. We focus on nuclear isotopes that correspond to the target nuclei of current and future experiments looking for WIMP- and neutrino-nucleus events. The nuclear structure calculations, performed in the context of the deformed shell model, are based on Hartree-Fock intrinsic states with angular momentum projection and band mixing for both the elastic and the inelastic channels.  Our predictions in the high-recoil-energy tail, show that  detectable distortions of the measured/expected signal may be interpreted through the inclusion of the non-negligible incoherent channels. 
\end{abstract}

\maketitle


\section{Introduction}

The recent observation of a coherent elastic neutrino nucleus scattering (CE$\nu$NS) process in the COHERENT experiment~\cite{Akimov:2017ade,Akimov:2020pdx} has opened up a wide range of new opportunities to test the standard model (SM) predictions~\cite{Bednyakov:2018mjd, Papoulias:2019lfi} as well as to investigate possible  new physics signatures~\cite{Papoulias:2018uzy} (for a recent review see Ref.~\cite{Papoulias:2019xaw}). Being a rapidly developing field, currently there are numerous projects aiming to measure this process at the Spallation Neutron Source (SNS)~\cite{Akimov:2018ghi}, at the European Spallation Source (ESS)~\cite{Baxter:2019mcx}, or near nuclear reactors such as CONUS~\cite{Hakenmuller:2019ecb}, CONNIE~\cite{Aguilar-Arevalo:2019jlr}, MINER~\cite{Agnolet:2016zir}, TEXONO~\cite{Wong:2010zzc}, RED100~\cite{Akimov:2016xdc}, RICOCHET~\cite{Billard:2016giu} and NUCLEUS~\cite{Angloher:2019flc}.  Furthermore, there is an intimate connection between dark matter (DM) and neutrino experiments~\cite{Moreno:2016hrs,Dror:2019onn,Dror:2019dib}.  Weakly interacting massive particles (WIMPs) are probably the most promising nonbaryonic cold DM candidates~\cite{Kortelainen:2006rd}. The latter arise in various frameworks beyond the SM and various experiments~\cite{Jungman:1995df} such as DarkSide~\cite{Agnes:2018oej}, DEAP-3600~\cite{Amaudruz:2017ekt}, CDEX~\cite{Yang:2019lao,Liu:2019kzq}, SuperCDMS~\cite{Agnese:2016cpb}, LUX~\cite{Akerib:2016vxi}, XENON1T~\cite{Aprile:2015uzo}, DARWIN~\cite{Aalbers:2016jon}, and PandaX-II~\cite{Fu:2016ega} are looking for tiny WIMP signals. The most appealing WIMP candidate is
the lightest supersymmetric particle (LSP) which is expected to be stable and
interacts very weakly with matter. In most cases, it is the lightest neutralino
which is a linear combination of the four neutral
fermions $\tilde{B}$, $\tilde{W_3}$, $\tilde{H_1}$, and $\tilde{H_2}$, with
$\tilde{B}$ and $\tilde{W_3}$ being the  supersymmetric (SUSY) partners of the
$U(1)$ gauge field $B$ and the third  component of the SU(2) gauge field
$W_3$, while $\tilde{H_1}$ and $\tilde{H_2}$ are the SUSY
partners of the light and heavy Higgs scalars~\cite{Kosmas:1997jm}.

The early data from the Cosmic
Background Explorer (COBE)~\cite{Smoot:1992td}  and Supernova Cosmology project~\cite{Gawiser:1998zh} implied that most of the DM is cold. 
Moreover, the recent WMAP~\cite{Hinshaw:2012aka} and Planck satellite~\cite{Aghanim:2018eyx} data showed that about 26.8\% mass of the universe is non-luminous
DM; the luminous matter forms only 4.9\% of the mass, with the other 68.3\%
being dark energy.
There are many experimental efforts to detect the elusive, yet to be observed
WIMPs which interact weakly with matter. Occasionally,
the latter will collide with the nuclei of the detector and the
resulting recoil may provide the finger-prints regarding their existence.
One such effort is the Super CDMS SNOLAB project (Sudbury, Canada) using  silicon and germanium
crystals.
For details regarding other experimental attempts,
see Refs.~\cite{Akerib:2016vxi,Fu:2016ega} and~\cite{Freese:2012xd, Liu:2017drf, Sahu:2017czz, Aprile:2018dbl, Amole:2017dex, Broniatowski:2009qu}. 
There are also efforts to search DM Axion candidates using low-noise
superconducting quantum amplifiers, being hypothetical particles
originally postulated to solve the strong CP problem. The Axion Dark Matter
Experiment (ADMX), at the University of Washington, reported results~\cite{Du:2018uak} showing that ``it is the world's first and only experiment to
have achieved the sensitivity'' to hunt for the DM axions. 

The odd-$A$ isotopes $^{127}$I, $^{133}$Cs and $^{133}$Xe, studied in this work, are among the most popular nuclei, used in many experiments as detectors, for both WIMP-nucleus~\cite{2012PhRvL.108r1301K, Bernabei:2013xsa, Behnke:2012ys,Divari:2000dc}  and \cevns searches~\cite{Akimov:2019wtg}. It is noteworthy that the detection techniques for WIMP-nucleus scattering and CE$\nu$NS are closely related, i.e., in both cases only nuclear recoil events are required. On the other hand, a \cevns event can perfectly mimic WIMP-nucleus scattering, thus being an irreducible background to direct detection DM searches, hence the neutrino floor is an important source of background~\cite{Billard:2013qya,OHare:2016pjy,Boehm:2018sux,OHare:2020lva}. This work is an extension of Ref.~\cite{Papoulias:2018uzy}, where we have investigated the impact of new physics interactions due to neutrino magnetic moments and $Z^\prime$ mediators by evaluating the expected CE$\nu$NS event rates in ton-scale direct DM detection experiments. According to our findings, the latter processes could potentially constitute a major source of background events. Specifically, these results indicate that the novel contributions may lead to a distortion of the expected recoil spectrum that could limit the sensitivity of direct search experiments.  At the same time, given the low-threshold capabilities of the next-generation DM experiments, the observation of solar \cevns events at a DM experiment will be a direct confirmation of its low mass WIMP sensitivity~\cite{AristizabalSierra:2019ykk}. While the potential of probing physics beyond the SM through \cevns measurements with large-scale DM detectors has been exhaustively tested, nuclear physics inputs remain a large source of uncertainty~\cite{AristizabalSierra:2019zmy}. Regarding elastic processes, the latter is encoded in the nuclear form factor, which if not properly addressed can lead to a misleading interpretation of a neutrino- or WIMP-induced signal. 

In the literature there are a considerable number of theoretical calculations which describe several aspects of \cevns and
direct detection of DM through nuclear recoils. For the case of CE$\nu$NS, realistic nuclear structure calculations have been previously performed, based on various models such as the quasiparticle random phase approximation (QRPA)~\cite{Papoulias:2015vxa}, microscopic quasiparticle phonon model (MQPM)~\cite{Pirinen:2018gsd}, coupled-cluster theory~\cite{Payne:2019wvy}, Hartree-Fock (HF) + Bardeen-Cooper-Schrieffer (BCS) calculations~\cite{Co:2020gwl},  and Hartree-Fock with a Skyrme (SkE2) nuclear potential~\cite{1805471}. For elastic WIMP-nucleus scattering, apart from the dominant scalar interaction, one needs to consider spin-spin
interaction coming from the axial current. On the other hand, for the case of inelastic scattering, the scalar interaction practically does not contribute.
The scalar interaction can arise from squark exchange, Higgs exchange,
the interaction of WIMPs with gluons, etc.
Some theoretical calculations are described in Refs.~\cite{Vergados:1999sf,
Holmlund:2004rv, Toivanen:2009zza, Pirinen:2016pxr,
Pirinen:2016pxr, Toivanen:2009zza}, while recently  in Ref.~\cite{Vergados:2016niz} the authors examined the possibility 
of detecting electrons in light WIMP mass searches. They considered a particle model involving WIMPs interacting with
electrons through the exchange of $Z$ bosons and found that event rates of 0.5--2.5 per kg.yr would be possible in this scenario. Moreover, shell model calculations~\cite{Menendez:2012tm, Klos:2013rwa, Baudis:2013bba, Vietze:2014vsa} have been performed to study DM event rates with $^{129,131}$Xe, $^{127}$I, $^{73}$Ge, $^{29}$Si, $^{27}$Al, $^{23}$Na, and $^{19}$F as detectors (for heavier nuclei, a truncated shell model space is employed).

In recent years, the deformed shell model (DSM), based on HF
deformed intrinsic states with angular momentum projection and band mixing,
has been established for the reliable description of several nuclear properties. The model proved quite successful in the mass range $A=$60--90~\cite{ks-book}, in describing
spectroscopic properties including spectroscopy of $N=Z$ odd-odd nuclei with
isospin projection~\cite{Srivastava:2014noa}, in double-beta-decay
half-lives~\cite{Sahu:2013yna, Sahu:2014nga}, in $\mu \to e$ conversion in the 
field of nuclei~\cite{Kosmas:2003xr} etc.
Recently it was employed to study the event rates for WIMP with $^{73}$Ge as the
detector~\cite{Sahu:2017czz}. In addition to the energy 
spectra and magnetic moments, the model was used to calculate the spin structure
functions, nuclear structure factors for elastic and inelastic WIMP-nucleus scattering. Furthermore, within the framework of the
DSM  we have recently calculated various new physics processes that could potentially contribute to \cevns and the respective neutrino-floor~\cite{Papoulias:2018uzy}, not only for $^{73}$Ge but also for other promising nuclei such as $^{71}$Ga,
$^{75}$As, and $^{127}$I. 

One of the purposes of the present paper is to
describe in detail the results for $^{127}$I, $^{133}$Cs, and $^{133}$Xe
and also to study the coherent and incoherent event rates for both neutrino-nucleus and WIMP-nucleus
scattering. We should stress that the DSM is applied to nuclei beyond $A$=90 for the first time in the present analysis. For
neutrino-nucleus scattering,  in addition to the aforementioned isotopes we have also  considered the $^{23}$Na, $^{40}$Ar, and $^{73}$Ge nuclear targets (DSM results for the spectroscopic properties of $^{23}$Na and $^{40}$Ar will be presented elsewhere).
The paper is organized as follows: in Sec.~\ref{sec:DSM} we summarize the most important features of the adopted nuclear DSM, while in Sec.~\ref{sec:DM} and Sec.~\ref{sec:neutrino-nucleus} we discuss the formalism of coherent and incoherent WIMP-nucleus and neutrino-nucleus scattering, respectively. Finally, our results are discussed in Sec.~\ref{sec:results} and our main conclusions in Sec.~\ref{sec:conclusions}. 

\section{Deformed shell model}
\label{sec:DSM}

The details of this model have  been described in our earlier
publications (for details see~\cite{ks-book}). Assuming axial symmetry for a given
nucleus, starting with a model space consisting of a given set of single-particle (sp) orbitals and an effective two-body Hamiltonian (TBME + spe), the
lowest-energy intrinsic states are obtained by solving the HF
single-particle equation self-consistently.  Excited
intrinsic configurations are obtained by making particle-hole excitations over
the lowest intrinsic state.  It is noteworthy that these intrinsic 
states denoted $\chi_K(\eta)$ do not have
definite angular momenta.  Hence, states of good  angular momentum, projected from
an intrinsic state $\chi_K(\eta)$, can be  written in the form
\begin{equation}
| \psi^J_{MK}(\eta) \rangle = \frac{2J+1}{8\pi^2\sqrt{N_{JK}}}\int d\Omega D^{J^*}_{MK}(\Omega)R(\Omega)| \chi_K(\eta) \rangle \, ,
\label{eqn.24}
\end{equation}
where $N_{JK}$ is the normalization constant given by
\begin{equation}
N_{JK} = \frac{2J+1}{2} \int^\pi_0 d\beta \sin \beta d^J_{KK}(\beta)\langle \chi_K(\eta)|e^{-i\beta J_y}|\chi_K(\eta) \rangle  \, .
\label{eqn.25}
\end{equation}
Here, $R(\Omega)=\exp(-i \alpha  J_z) \exp(-i \beta J_y) \exp(-i \gamma J_z)$ denotes the general rotation operator and $\Omega$ represents the Euler angles ($\alpha$, $\beta$,
$\gamma$).  The good
angular momentum states projected from different intrinsic states are not in
general orthogonal to each other, hence, band mixing calculations are performed after appropriate orthonormalization.  The resulting eigenfunctions are of 
the form
\begin{equation}
\vert\Phi^J_M(\eta) \rangle \, =\,\sum_{K,\alpha} S^J_{K \eta}(\alpha)\vert 
\psi^J_{M K}(\alpha)\rangle \, ,
\label{phijm}
\end{equation}
with  $S^J_{K \eta}(\alpha)$ being the expansion coefficients.
The nuclear matrix elements occurring in the calculation of magnetic moments,
elastic and inelastic spin structure functions etc. are evaluated using 
the wave functions 
$| \Phi^J_M(\eta) \rangle$. DSM is well established to be a 
successful model for transitional nuclei (with $A$=60--90)~\cite{ks-book,Srivastava:2014noa,Sahu:2013yna,Sahu:2002vc,Sahu:2003xf}. 

\begin{figure}[t]
\centering
\includegraphics[width=0.49\linewidth]{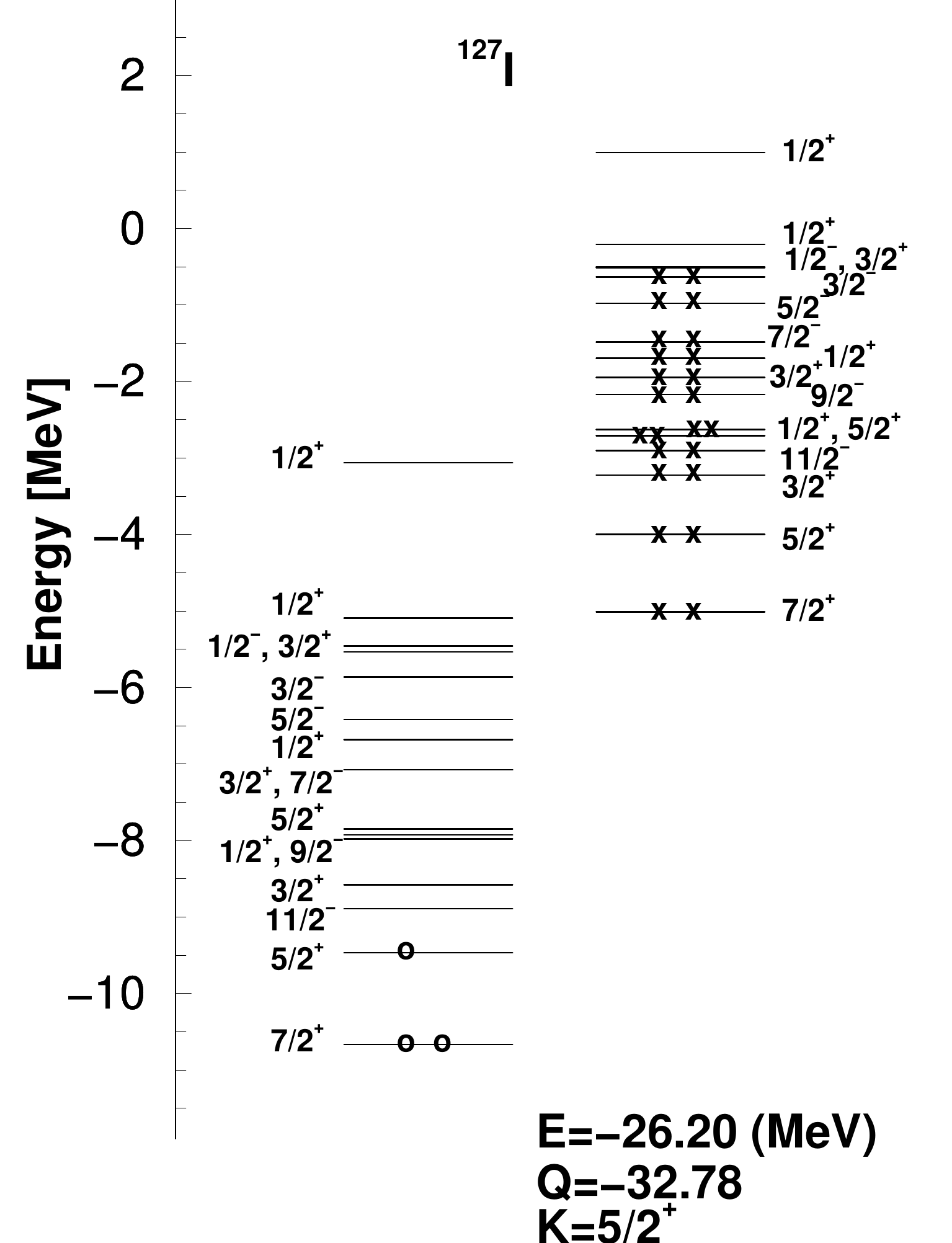}
\includegraphics[width=0.49\linewidth]{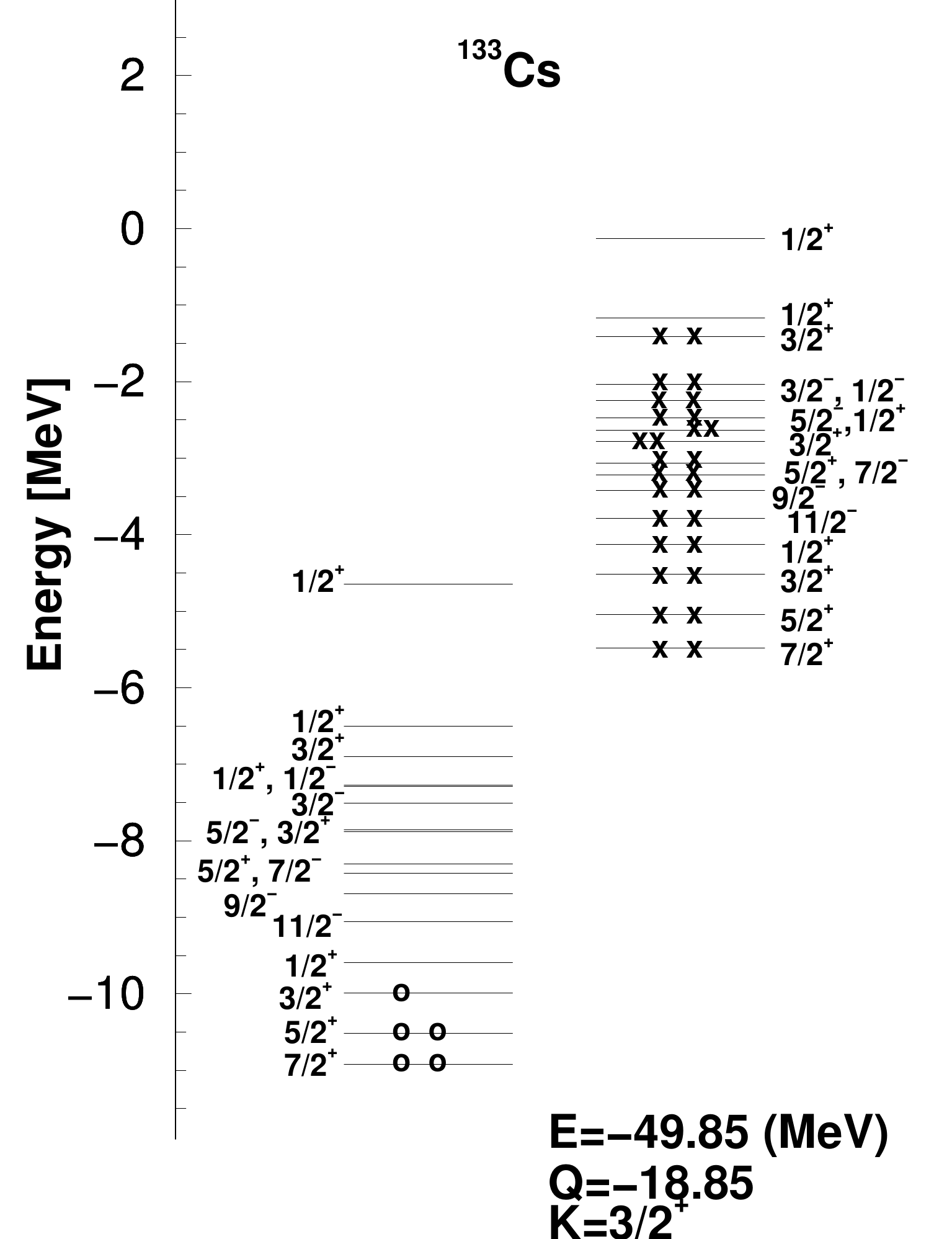}
\includegraphics[width=0.49\linewidth]{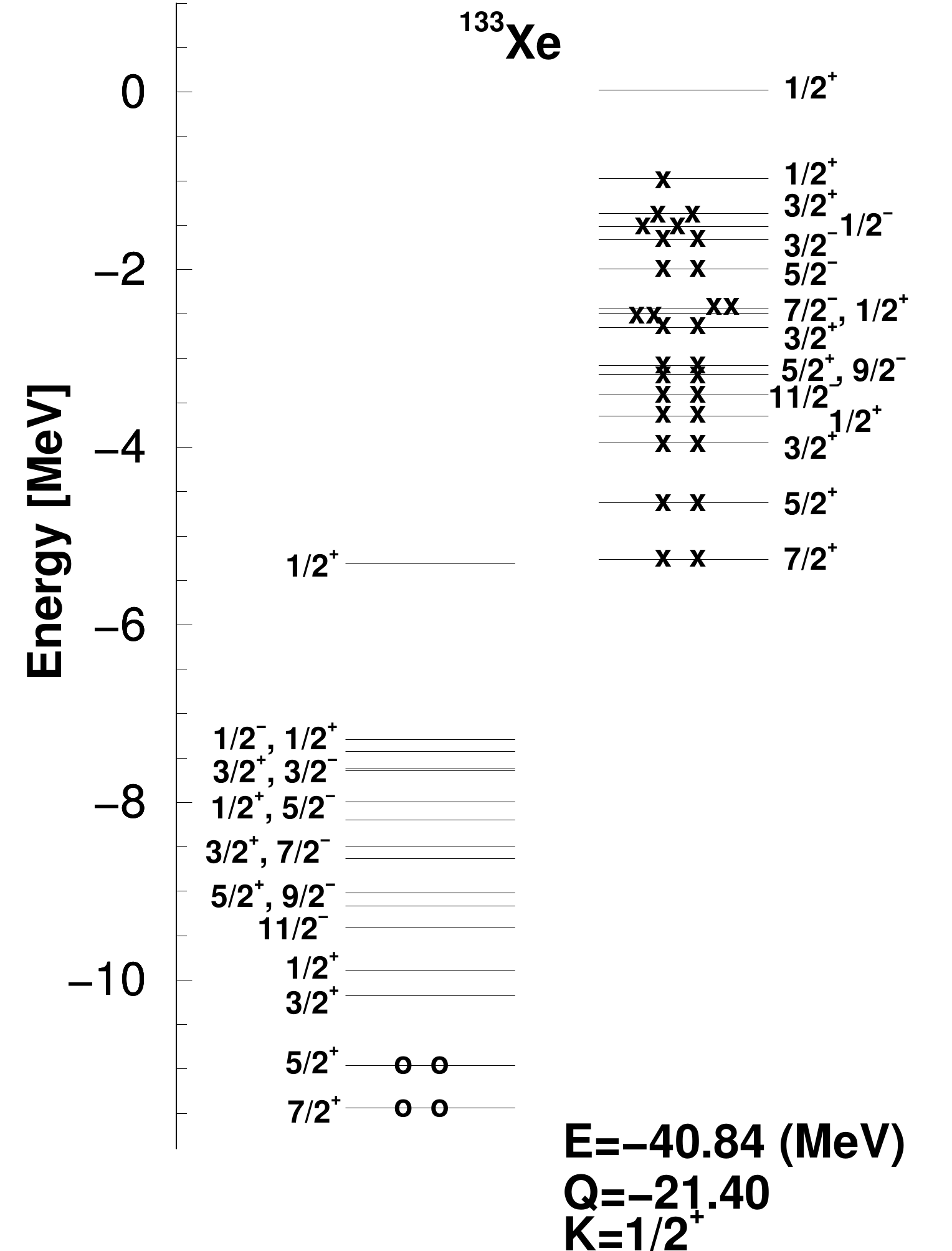}
\caption{HF single-particle energy spectra for $^{127}$I, $^{133}$Cs, and 
$^{133}$Xe. Circles  represent protons  and {\small $\boldsymbol{\times}$}'s represent neutrons. The HF energy $E$ (in MeV), mass quadrupole moment $\mathsf{Q}$ (in units of the square of the oscillator length parameter), and the total azimuthal quantum number $K$ are also shown.}
\label{fig:hf}
\end{figure}

\begin{figure}[t]
\centering
\includegraphics[width=0.49\linewidth]{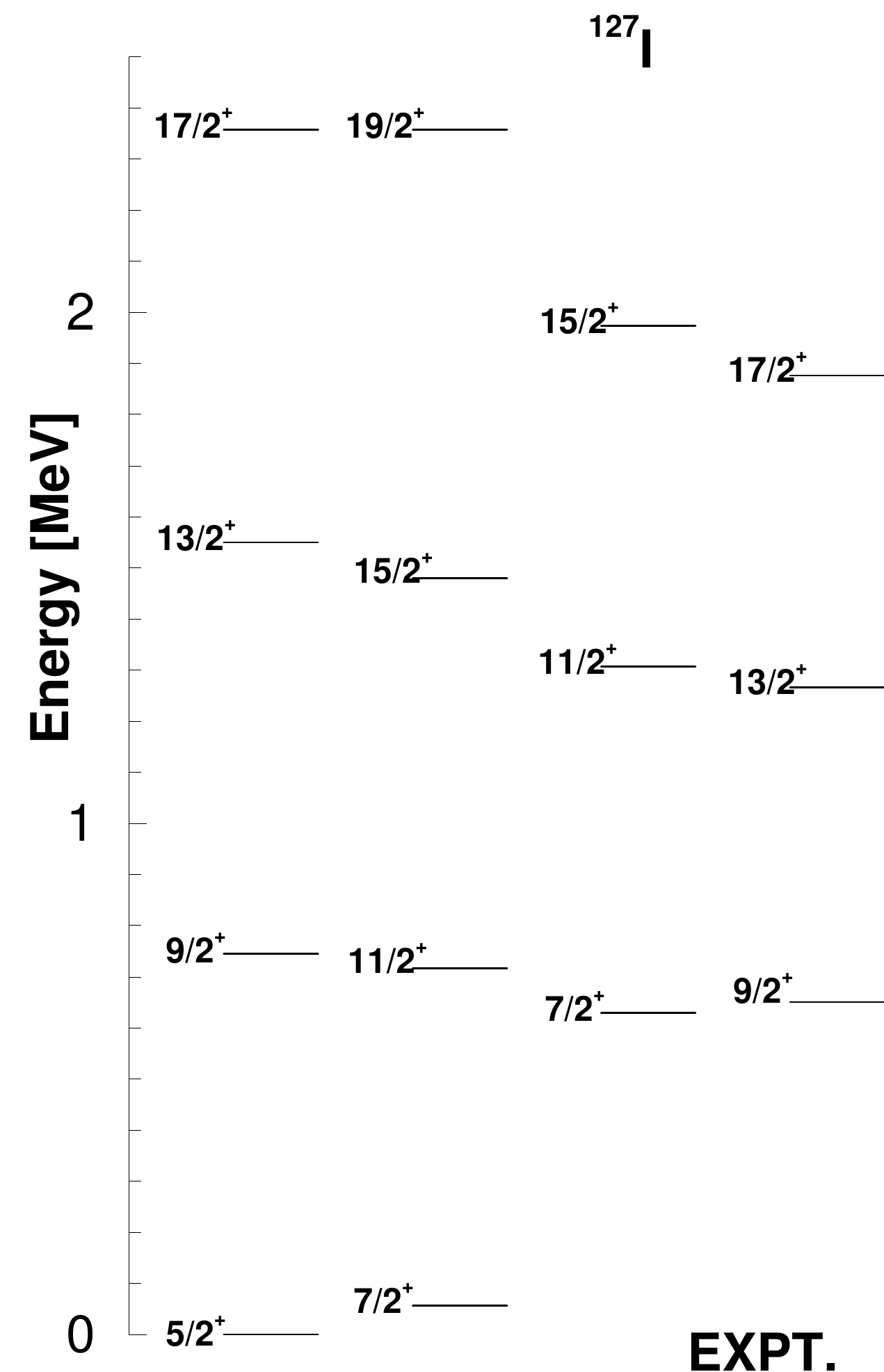}
\includegraphics[width=0.49\linewidth]{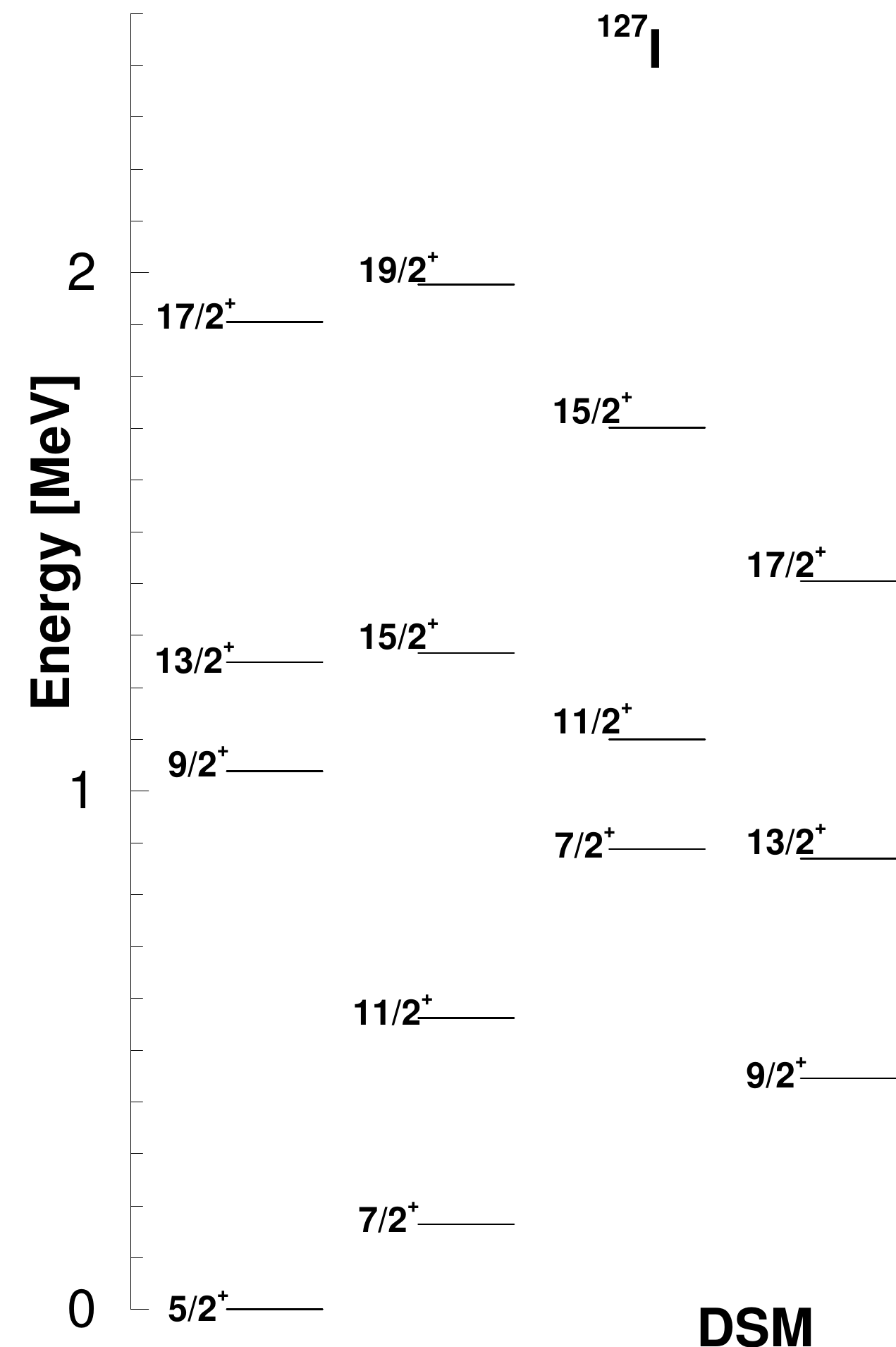}
\caption{Comparison of the DSM energy spectrum with the experimental data for $^{127}$I. 	Experimental values are taken from~\cite{nndc}.}
\label{fig:spectra1}
\end{figure}

In order to evaluate the WIMP-nucleus scattering event rates, we first calculate the energy spectra and magnetic moments by employing the DSM wave functions and then  compare our results with available experimental data~\cite{nndc}. A good agreement with experiment ensures the reliability of our predictions on the event rates.
In our nuclear structure calculations for $^{127}$I, $^{133}$Cs, and $^{133}$Xe, the single-particle orbits, their energies, and the assumed effective interaction (obtained by re-normalizing the CD-Bonn potential) are taken from 
Ref.~\cite{Coraggio:2017bqn}. The model space
consists of the orbitals $0g_{7/2}$, $1d_{5/2}$, $1d_{3/2}$, $2s_{1/2}$,
and $0h_{11/2}$ with the closed core $^{100}$Sn. 
The single particle energies for the five orbitals are 0.0, 0.4,
1.4, 1.3, and 1.6 MeV for protons and 0.0, 0.7, 2.1, 1.9, and 3.0 MeV for
neutrons as given in~\cite{Coraggio:2017bqn}, while we neglect the Coulomb term in our calculations.
From the experimentally measured static quadrupole moment $\mathsf{Q}$ for the $^{127}$I nucleus, the experimental negative sign of the $E2/M1$ mixing ratios for the lighter iodine isotopes, and also different theoretical analyses, Ding \emph{et al.}~\cite{Ding:2012zzb} concluded that $^{127}$I has a slightly oblate shape at least at low excitation. Hence we restrict ourselves to HF solutions with an oblate shape for this nucleus.
The lowest oblate HF single particle spectrum is shown in Fig.~\ref{fig:hf}.
Also, for $^{133}$Cs and $^{133}$Xe the lowest HF intrinsic state, which is oblate,
is also shown in the same figure. As described
before, we obtained the lowest HF configuration by performing an axially 
symmetric HF calculation for each nucleus. Then various
excited configurations were obtained by making particle-hole excitations
over this lowest HF configuration. We have 
chosen a total of  six, three and four configurations for $^{127}$I, $^{133}$Cs, and 
$^{133}$Xe, respectively. These choices reproduce reliably
the spectroscopic properties of the above nuclear isotopes. 
The calculated energy levels obtained from angular momentum projection and band mixing
for each nucleus are classified into collective bands
on the basis of the $E2$ transition probabilities between them. As an example, the calculated
energy spectrum for $^{127}$I is compared with the available experimental
data in Fig.~\ref{fig:spectra1}.

For the $^{127}$I isotope, four low-lying
positive parity collective bands have been identified experimentally
\cite{Ding:2012zzb} and are reasonably well described within the DSM.
First, the DSM reproduces correctly the ground-state $5/2^+$
level. Second, experimentalists have suggested that the low-lying
positive-parity bands in this nucleus should be associated with the proton $d_{5/2}$ and
$g_{7/2}$ configurations. An analysis of the DSM
wave functions shows that these
collective bands originate either from the
lowest HF configuration given in Fig.~\ref{fig:hf}, where the odd proton is in 
the $k=5/2^+$ deformed orbit, or from the first excited 
intrinsic state, which mainly originates from the spherical $1d_{5/2}$ and 
$0g_{7/2}$ orbits. 
Thus, using the DSM we also predict that the collective bands for this 
nucleus
originate mainly from the $d_{5/2}$ and $g_{7/2}$ orbitals.
Turning briefly to $^{133}$Cs, it can be seen that the
DSM calculation
reproduces correctly the ground state, which is $J=7/2^+$. In addition, a collective band
built on this level has been observed for this nucleus and the calculated
$K=7/2^+$ band is found to be somewhat compressed compared to experiment. This collective band originates mainly from the
lowest oblate intrinsic state.
Finally, for the case of $^{133}$Xe, the experimental data show that well-defined collective bands are not present in this nucleus. The DSM is found to reproduce the ground-state spin ($3/2^+$) and also the ordering of the excited
levels (spin of the first excited state is $1/2^+$).
In conclusion, for the above three nuclei the DSM reproduces the spin of the ground
state and the first excited level correctly, and the energy of the first excited level reasonably well. Note that, these two levels are 
important for the study of the elastic and inelastic scattering processes we are undertaking.

Clearly the DSM gives results in agreement with available experimental data and very close to those obtained from full shell model calculations for physical observables such as energy spectra, and $B(E2)$ and $B(M1)$ values (see, e.g., Refs.~\cite{ks-book,Srivastava:2014noa}). As  is well known, $B(M1)$ values are related to the square of the off-diagonal matrix element of the $M1$ operator, whereas the magnetic moment connects diagonal matrix elements. Thus, from such applications to different nuclear systems of the DSM the confidence level acquired is high, provided that the effective interaction in the chosen model space is reasonable. In addition, through the choice of the intrinsic states of the DSM model, a satisfactory description of energy spectra for $^{127}$I, $^{133}$Cs, and $^{133}$Xe was obtained as shown in Fig.~\ref{fig:spectra1}.

\begin{table}
	\caption{Values of the harmonic oscillator parameter $b$ (in units of~fm) used for different nuclei.}
\begin{tabular}{c|cccccc}
\hline
Nucleus & $^{23}$Na  &  $^{40}$Ar   & $^{73}$Ge & $^{127}$I & $^{133}$Cs   &  $^{133}$Xe\\
\hline
$b$~(fm) & 1.573      &   1.725      &    1.90   &2.09    &   2.11       &    2.11  \\
\hline
\end{tabular}
\label{tab-2}
\end{table}

\begin{table}
\caption{Calculated DSM magnetic moments for the different nuclear states of 
        $^{127}$I, $^{133}$Cs, and $^{133}$Xe.
	Results refer to the bare gyro-magnetic ratios and are compared with the experimental ones from Ref.~\cite{nndc}.} 
\begin{tabular}{c|ccccccc}
\hline
       Nucleus      &$J$   & $<l_p>$ & $<S_p>$ & $<l_n>$ & $<S_n>$ & $\mu$ (nm) & Expt. \\
\hline
$^{127}$I   & $5/2^+$ &2.395    & -0.211     &0.313  & 0.002 & 1.207 & 2.813 \\
      & $7/2^+$     &2.580    & -0.243    & 1.097 & 0.007   &  0.969&2.54    \\
      & $3/2^+$ & 1.542   & -0.181     & 0.148 & -0.008   & 0.560 & 0.97 \\
      &         &         &            &       &          &       &      \\
$^{133}$Cs&$7/2^+$&3.40   & -0.34      &0.49   & -0.048   & 1.69  & 2.582 \\
          &$5/2^+$&2.47   & -0.25      &0.30   & -0.034   & 1.23  & 3.45 \\
      &         &         &            &       &          &       &      \\
$^{133}$Xe&$3/2^+$&0.39   & -0.04      &1.44   &-0.285    & 1.26  & 0.81 \\
          &$1/2^+$&0.30   & -0.03      &0.004  & 0.230    &-0.75  &      \\
\hline
\end{tabular}
\label{tab-1}
\end{table}

Regarding the values of the harmonic oscillator size parameter $b$  used for different nuclei, they have been chosen assuming the well-known formula ($b \propto A^{1/6}$) and they are listed in Table~\ref{tab-2}. 
In our earlier work~\cite{Kosmas:2003xr} in the calculation of transition matrix  elements regarding $\mu \to e$ conversion in $^{72}$Ge, we had taken the value $b=1.90$~fm. 
The latter, provided reasonably good agreement for proton and neutron form factors and coherent matrix elements compared to other calculations and also with available experimental data. The oscillator parameters for the three nuclei are taken in a similar manner. We stress that small variations of $b$ have no essential impact on the form factors specifically for the momentum transfer of interest in our present work.

We mention that in the calculation of WIMP-nucleus event rates, the nuclear spin plays an important role,  and   therefore,  we also calculate the nuclear magnetic moments for the isotopes in question. The calculated magnetic moments, predicted in the framework of the DSM, for  $^{127}$I, $^{133}$Cs, and $^{133}$Xe are listed in Table~\ref{tab-1} and compared with the experimental data~\cite{nndc}. The contribution of protons
and neutrons to the orbital and spin parts are also given in this table
for the reader's convenience.  Focusing on $^{127}$I, the magnetic moments for the 
levels $5/2^+$, $7/2^+$, and $3/2^+$ are experimentally known, while the DSM
calculated values with bare gyro-magnetic moments are off by about a factor of 2. 
A similar trend is also found in the case of the $^{133}$Cs isotope, while for $^{133}$Xe the calculated value is 50\% larger than the experimental one. 
It should be noted, however, that shell model calculations by Coraggio \emph{et al.}~\cite{Coraggio:2017bqn}, using  state-dependent effective
charges for the $B(E2)$ values and state-dependent quenching factors for Gamow-Teller transitions $GT^+$, obtained a much better agreement with the experimental data. The latter implies that a similar approach with state-dependent gyro-magnetic moments in our DSM calculations is expected to reproduce the experimental 
magnetic moments for the nuclei in question much better. 

Indeed, the use of effective gyromagnetic ratios  with  values $g_l=1.5$, $g_s^p=4$, and $g_s^n=-3.0$ leads to improved gyromagnetic moments, e.g., 2.742, 2.877, and 1.613 for the $J=5/2^+$, $7/2^+$, and $3/2^+$ states of $^{127}$I, respectively. For the states $7/2^+$ and $5/2^+$ in $^{133}$Cs, the new magnetic moments are 3.884 and 2.807. For $^{133}$Xe, the new magnetic moment is 1.280. These new values of magnetic moments are in better agreement with experiment. We note that these effective gyromagnetic moments are not unique and one can obtain similar values with new sets of effective gyromagnetic ratios. Many theoretical nuclear structure calculations use effective gyromagnetic ratios in the calculation of magnetic moments. The use of effective gyromagnetic ratios does not change the nuclear wave function and hence the calculated form factors, energy spectra, etc., remain unchanged.

We furthermore note 
that, for $^{71}$Ga, $^{73}$Ge, and $^{127}$I, Holmlund \emph{et al.}~\cite{Holmlund:2004rv} have calculated the magnetic moments
as well as their orbital and spin parts within the MQPM and compared
the obtained results with the predictions of other models. For $^{127}$I, they concluded that, even though MQMP predictions  were close to the single-particle 
estimates, the experimental values of the magnetic moments could not be exactly reproduced
quantitatively. Therefore, to achieve a better reproducibility of the experimental data, detailed shell model calculations are required.

\section{Event rates for WIMP-nucleus scattering}
\label{sec:DM}

The WIMP flux coming from the galactic halo on the Earth is expected to be quite
large, of the order of $10^5~\mathrm{cm^{-2}\, s^{-1}}$. Even though the interaction
of WIMP with matter is weak, this flux is sufficiently large for the
galactic WIMPs to deposit a measurable amount of energy in an appreciably
sensitive detector apparatus when they scatter off nuclei. Most of the
experimental searches of WIMPs are based on the
direct detection through their interaction with target nuclei.
The relevant formalism of WIMP-nucleus scattering  has been discussed in our
earlier works~\cite{Sahu:2017czz,Papoulias:2018uzy} (see also Refs.~\cite{Toivanen:2009zza,Kortelainen:2006rd,Holmlund:2004rv}).
For completeness,  however, below we summarize briefly  the  most important steps.
Here, we consider the spin-spin interaction and  the scalar
interaction. In the case of the former, the WIMP couples to the spin of the
nucleus, while the scalar interaction is proportional to the square of the mass number $A$.
In the expressions for the event rates, the particle physics (SUSY) part is
separated from the nuclear part so that the role played by the nuclear physics 
part becomes apparent.

\subsection{Wimp-nucleus elastic scattering}

The differential event rate per unit detector mass for a WIMP with mass
$m_\chi$ can be written as~\cite{Jungman:1995df}
\begin{equation}
dR = N_t\; \phi\; f \; \frac{d\sigma}{d \vert q \vert ^2} d^3 v\; 
d\vert q \vert^2 \, ,
\label{eqn.1}
\end{equation}
where $\phi=\rho_0 v/m_\chi$   
is the DM flux, with $\rho_0$ being the  local WIMP density, $N_t$ stands for the number of target nuclei per unit mass, and 
$f$ denotes the WIMP velocity distribution, 
which is assumed to be of the Maxwell-Boltzmann type.
The latter takes into account the distribution of the WIMP velocity relative to the detector (or Earth) and also the motion of the Sun and Earth. 
Note that when  the
rotation of Earth around its own axis is neglected, then $v=\vert {\mathbf v} \vert$ corresponds to the 
relative velocity of the WIMP with respect to the detector. Finally, $q \equiv |\mathbf{q}|$ represents the magnitude of the 
3-momentum  transfer to the nuclear target, which is related to the dimensionless
variable  $u=q^2b^2/2$m with $b$ denoting the harmonic oscillator length parameter.

The WIMP-nucleus differential cross section in the laboratory frame is then given by~\cite{Sahu:2017czz, Pirinen:2016pxr,Toivanen:2009zza,Kortelainen:2006rd,Holmlund:2004rv}
\begin{equation}
\frac{d\sigma (u,v)}{du} = \frac{1}{2}\, \sigma_0\,\left(\frac{1}{m_pb}
\right)^2 \,\frac{c^2}{v^2} \,\frac{d\sigma_{A}(u)}{du} \, ,
\label{eqn.2}
\end{equation}
with 
\begin{equation}
\begin{aligned}
\dis\frac{d\sigma_{A}(u)}{du}  =  & (f_A^0)^2 F_{00}(u) +
2f_A^0 f_A^1  F_{01}(u) + (f_A^1)^2 F_{11}(u) \\   & +
\left[ Z \l(f_S^0 + f_S^1 \r) \right]^2 |F_Z(u)|^2 \\& + \left[ (A-Z) \l(f_S^0 - f_S^1 \r) \right]^2 |F_N(u)|^2 \\ & + 2 Z (A-Z) \left[(f_S^0)^2 - (f_S^1)^2 \right] |F_Z(u)||F_N(u)| \, ,
\label{eqn.3}
\end{aligned}
\end{equation}
where $F_Z(u)$ and $F_N(u)$ denote the nuclear form factors for protons and neutrons respectively.
In the latter expression, the first three terms correspond to the spin contribution,
coming from the axial current, and the last three terms account for the coherent
part, coming mainly from the scalar interaction. 
Here, $f_A^0$ and $f_A^1$ represent isoscalar and isovector parts of the axial
vector current, and similarly  $f_S^0$ and $f_S^1$ represent isoscalar and
isovector parts of the scalar current. The nucleonic current parameters  $f_A^0$
and $f_A^1$ depend on the specific SUSY model assumed in this work for the WIMP (LSP). However, $f_S^0$ and
$f_S^1$ depend  on the hadron model used to embed 
quarks and gluons into nucleons~\cite{Vergados:1995hs}. 

The normalized spin structure functions $F_{\rho\rho'}(u)$ with
$\rho$, $\rho'$ = 0,1 in Eq. (\ref{eqn.3}) are defined as
\begin{equation}
F_{\rho\rho'}(u) = \dis\sum_{\lambda,\kappa}\frac{\Omega_\rho^
{(\lambda,\kappa)}(u)\Omega_{\rho'}^{(\lambda,\kappa)}(u)}{\Omega_\rho
\Omega_{\rho'}}\, ,
\end{equation}
\begin{equation}
\Omega_\rho^{(\lambda,\kappa)}(u) = \sqrt{\frac{4\pi}{2J_i + 1}} \\
                \times \langle J_f \| \dis\sum_{j=1}^A \left[Y_\lambda(\Omega_j)
\otimes \sigma(j)\right]_\kappa j_\lambda(\sqrt{u}\,r_j) 
\omega_\rho(j) \|J_i\rangle \, ,
\label{eqn.6}
\end{equation}
where $\omega_0(j)=1$ and $\omega_1(j)=\tau(j)$ with $\tau=+1$ for protons and $\tau=-1$ for neutrons, $j_\lambda$ is the spherical Bessel function, and the static spin matrix elements are defined as $\Omega_\rho(0) =
\Omega_\rho^{(0,1)}(0)$. The  reduced matrix element appearing in Eq. (\ref{eqn.6})   is then
evaluated within the framework of the DSM. For this purpose, we need the sp 
matrix elements of 
the operator of the form $t^{(l,s)J}_\nu$, given by
\be
\barr{l}
\langle n_il_ij_i\| \hat{t}^{(l,s)J}\| n_kl_kj_k \rangle = \\
\\
\dis\sqrt{(2j_k+1)(2j_i+1)(2J+1)(s+1)(s+2)} \\
\\
\left\{\begin{array}{ccc}
l_i & 1/2 & j_i\\
l_k & 1/2 & j_k\\
l   & s   & J
\end{array}\right\} \;
\langle l_i \| \sqrt{4\pi}Y^l \|l_k \rangle \;\langle n_il_i |j_l(kr)| 
n_kl_k \rangle \, ,
\earr \label{eqn.26}
\ee
where $\{\}$ denotes the 9-$j$ symbol. 
Assuming that  the polar axis is aligned along the
direction of ${\mathbf v}_1$ (velocity of the Earth with respect to the Sun)
and converting the integration variables into 
dimensionless form, the event rate is obtained by integrating
Eq. (\ref{eqn.1}) with respect to $u$, the velocity $v$, and the scattering angle $\theta$, as ~\cite{Sahu:2017czz} 
\begin{equation}
\langle R \rangle_{\text{el}} = \int^1_{-1} d\xi  \int^{\psi_{max}}_{\psi_{min}} d\psi 
\int^{u_{max}}_{u_{min}}
G(\psi, \xi) \frac{d\sigma_{A}(u)}{du} du \, .
\label{eqn.9}
\end{equation}
In the above expression, $G(\psi, \xi)$ is given by
\begin{equation}
G(\psi, \xi) = \frac{\rho_0}{m_\chi} \frac{\sigma_0}{Am_p} \left(\frac{1}
{m_pb}\right)^2 \frac{c^2}{\sqrt{\pi}v_0} \psi e^{-\lambda^2} e^{-\psi^2}
e^{-2\lambda\psi\xi} \, ,
\label{eqn.10}
\end{equation}
with $\psi=v/v_0$, $\lambda=v_E/v_0$, $\xi=\cos(\theta)$.
For our calculation we employed the following parameters: WIMP density
$\rho_0 = 0.3 \;\mathrm{GeV/{cm^3}}$, $\sigma_0 = 0.77 \times 10^{-38} \mathrm{cm^2}$, and proton mass $m_p = 1.67 \times 10^{-27}$~kg. The
velocity of the Sun with respect to the galactic center is taken to be
$v_0 =220$ km/s and the velocity of the Earth relative to the Sun is taken as
$v_1=30$ km/s. The velocity of the Earth with respect to the galactic
center $v_E$ is given by
$v_E = \sqrt{v_0^2 + v_1^2 + 2v_0v_1 \sin(\gamma) \cos(\alpha)}$,
where $\alpha$ is the modulation angle, which stands for the phase of the Earth on  its
orbit around the Sun, and $\gamma$ is the angle between the normal to the
ecliptic and the galactic equator~\cite{Kosmas:1997jm}, which is taken to be $\simeq 29.8^\circ$. 

For simplicity, by writing  
$X(1)=  F_{00}(u)$, 
$X(2)= F_{01}(u)$,
$X(3)=   F_{11}(u)$,
$X(4) = |F_Z(u)|^2$,
$X(5) = |F_N(u)|^2$,
$X(6) = |F_Z(u)||F_N(u)|$,
the event rate per unit mass of the detector of Eq. (\ref{eqn.9}) can be cast in the form
\begin{equation}
\begin{aligned}
\langle R \rangle_\text{el}  =&  (f^0_1)^2 D_1 + 2 f^0_Af^1_A D_2 + (f^1_A)^2 D_3  \\ 
& + \left[Z \l(f_S^0 + f_S^1 \r) \right]^2 D_4 \\ & +  \left[(A-Z) \l(f_S^0 - f_S^1 \r) \right]^2 D_5 \\ & + 2 Z (A-Z) \l[(f_S^0)^2 - (f_S^1)^2 \r] D_6 \, ,
\end{aligned}
\label{eqn.12}
\end{equation}
where $D_i$  is the three dimensional integrations of Eq. (\ref{eqn.9}), defined as
\begin{equation}
D_i = \int^1_{-1} d\xi  \int^{\psi_{\text{max}}}_{\psi_{\text{min}}} d\psi 
\int^{u_{\text{max}}}_{u_{\text{min}}}
G(\psi, \xi) X(i) \, du  \, , \quad i=1,\ldots,6 \, .
\label{eqn.9a}
\end{equation}
The lower and upper limits of integrations given in Eqs. (\ref{eqn.9}) and 
(\ref{eqn.9a}) are taken from Ref.~\cite{Pirinen:2016pxr} and read
\begin{equation}
\begin{aligned}
\psi_{\text{min}} =& \frac{c}{v_0} \left(\frac{Am_p T_\mathrm{thres} }{2\mu^2_r}\right )^{1/2} \, , \\
\psi_{\text{max}} =& -\lambda\xi + \sqrt{\lambda^2\xi^2+\frac{v_{esc}^2}{v_0^2} -1
- \frac{v^2_1}{v^2_0}-\frac{2 v_1}{v_0} \sin(\gamma) \cos(\alpha)} \, ,\\
u_{\text{min}} = & Am_p T_\mathrm{thres} b^2\, , \\
u_{\text{max}}= & 2(\psi\mu_rbv_0/c)^2 \, .
\end{aligned}
\end{equation}
Assuming the escape velocity from our galaxy to be $v_{esc}=625$ km/s, the
quantity $v_{esc}^2/v_0^2 -1- v^2_1/v^2_0$ appearing in the definition of $\psi_\text{max}$ is equal to $7.0525$, and  similarly $(2 v_1/v_0) \sin(\gamma)=0.135$,
while, $T_\mathrm{thres}$ denotes the detector threshold energy and $\mu_r$ is 
the reduced mass of the WIMP-nucleus system.

\subsection{WIMP-nucleus inelastic scattering}

In the case of inelastic WIMP-nucleus scattering, the  initial and final nuclear states do not coincide and the corresponding cross section due to the scalar current is considerably 
smaller with respect to the elastic case.
 We then focus on the spin dependent scattering and the  
 inelastic event rate per unit mass of the detector can be written as
\begin{equation}
\langle R \rangle_{\text{inel}} = (f^0_1)^2 E_1 + 2 f^0_Af^1_A E_2 + (f^1_A)^2 E_3 \, ,
\label{eqn.19}
\end{equation}
where $E_1$, $E_2$ and $E_3$ are the three dimensional integrals
\begin{equation}
E_i = \int^1_{-1} d\xi  \int^{\psi_{\text{max}}}_{\psi_{\text{min}}} d\psi 
\int^{u_{\text{max}}}_{u_{\text{min}}}
G(\psi, \xi) X(i) \, du \, .
\label{eqn.20}
\end{equation}
The integration limits in the latter expression read~\cite{Pirinen:2016pxr, 
Toivanen:2009zza}
\begin{equation}
u_{\text{min}} = \frac{1}{2}b^2\mu_r^2\frac{v^2_0}{c^2}\psi^2
\left[ 1 - \sqrt{1-\frac{\Gamma}{\psi^2}} \right ]^2 \, ,
\label{eqn.21}
\end{equation}
\begin{equation}
u_{\text{max}} = \frac{1}{2}b^2\mu_r^2\frac{v^2_0}{c^2}\psi^2
\left[ 1 + \sqrt{1-\frac{\Gamma}{\psi^2}} \right ]^2 \, ,
\label{22}
\end{equation}
where 
\begin{equation}
\Gamma = \frac{2 E^*}{\mu_rc^2} \frac{c^2}{v_0^2} \, ,
\label{23}
\end{equation}
with $E^*$ being the energy of the excited nuclear state. Here, $\psi_{\text{max}}$ is the same as in the
elastic case and the lower limit  $\psi_{\text{min}} = \sqrt{\Gamma}$. The rest of the 
parameters, e.g., $\rho_0$, $\sigma_0$, etc., take the same values as in the 
elastic case.

\section{neutrino-nucleus scattering}
\label{sec:neutrino-nucleus}

In this section we consider neutrino-nucleus scattering focusing on both coherent ($g.s. \rightarrow g.s.$ transitions) and incoherent channels  ($g.s. \rightarrow $\emph{excited-state} transitions). In the low-energy regime of SM electroweak interactions, the latter can be well studied through an effective Lagrangian written in the approximation of four-fermion contact interaction  and normalized to the Fermi coupling constant $G_F$, as
\begin{equation}
\mathcal{L}(x) = \frac{G_F}{\sqrt{2}} L_\mu(x)H^\mu(x) \, ,
\label{eq:lepton_current}
\end{equation}
where the corresponding leptonic and hadronic currents are given  through the expressions 
\begin{equation}
\begin{aligned}
L_\mu(x) = &  \overline{\nu}(x)\gamma_\mu(1-\gamma_5)  \nu(x) \, ,\\
H^\mu(x) = & \sum_{f=n,p}  \overline{\psi}_f(x)\gamma^\mu\left(g_V^f-g_A^f\gamma_5\right)\psi_f(x)  \, .
\end{aligned}
\label{eq:hadronic_field_current}
\end{equation}
The left- and right-handed couplings are written in terms of the usual vector (axial-vector) $g_V^{f}$ ($g_A^{f}$) couplings as functions of the weak mixing angle $\sin^2 \theta_W \equiv s_W^2=0.2386$
\begin{equation}
g_L^{f}  =\frac{1}{2}\left(g_V^{f}+g_A^{f}\right), \quad 
g_R^{f}  =\frac{1}{2}\left(g_V^{f}-g_A^{f}\right) \, ,
\label{eq:left_right_couplings}
\end{equation}
where the fundamental SM couplings of the nucleon $f=\{n,p\}$ to the $Z_0$ boson read

\begin{eqnarray}
\begin{aligned}
g_V^p   &=   \frac{1}{2}-2 s_W^2,   & g_A^p &=  \frac{1}{2}   \\
g_L^p   &=    \frac{1}{2}\left(1-2s_W^2\right),    & g_R^p &=   -s_W^2 \\ 
g_V^n   &=  -\frac{1}{2},   & g_A^n  &= -\frac{1}{2} \\
g_L^n   &= -1,     & g_R^n  &= 0 \, .\\
\end{aligned}
\label{tab:sm_couplings}
\end{eqnarray}

In what follows, the neutrino-nucleus cross sections and the corresponding event rates rely on the formalism of Bednyakov and Naumov (BN)~\cite{Bednyakov:2018mjd}, while the nuclear physics effects are incorporated  by employing the DSM formalism described in Sec.~\ref{sec:DSM}. A key difference with respect to Ref.~\cite{Bednyakov:2018mjd} is that the DSM method adopted here predicts the excitation energies of the nuclei in question, in contrast to the BN study, where the latter are put by hand. We furthermore provide differential and integrated event rates expected in various nuclear targets for different neutrino sources.

 To set up the notation, we provide below the most important features of the BN formalism.  We start by denoting the 4-momentum of incoming and outgoing neutrinos  as $k=(E_\nu,\mathbf{k})$ and $k'=(E_\nu',\mathbf{k}')$, respectively, and the 4-momentum of the initial (final) nuclear state as $P_n$  ($P'_m$). Then, the energy of the outgoing neutrino is written in terms of the incident neutrino energy, the scattering angle $\theta$ between $\mathbf{k}$ and $\mathbf{k}'$, the nuclear excitation energy $\Delta\varepsilon_{mn}$ (i.e., the energy difference between the initial and the final nuclear states), and the nuclear mass $m_A$, as~\cite{Bednyakov:2018mjd}
\begin{equation}
\label{eq:outgoing_neutrino_energy_lab}
E_\nu' = \frac{m_A(E_\nu-\Delta\varepsilon_{mn})-E_\nu\Delta\varepsilon_{mn}+\Delta\varepsilon_{mn}^2/2}{m_A+E_\nu(1-\cos\theta)-\Delta\varepsilon_{mn}}\, ,
\end{equation}
while the energy of the recoiling nucleus reads
\begin{equation}
T_A = \sqrt{m_A^2+\mathbf{q}^2}-m_A \, ,
\label{eq:recoil}
\end{equation}
where $\mathbf{q}$ is the 3-momentum transfer with magnitude
\begin{equation}
q \equiv |\mathbf{q}|  = \left( E_\nu^2+E'^2_\nu-2E_\nu E_\nu'\cos\theta \right)^{1/2} \simeq (2 m_A T_A)^{1/2} \, .
\end{equation}
Finally, it is noteworthy that the nuclear recoil energy given in Eq. (\ref{eq:recoil}) can be adequately approximated by keeping only the first term in the $1/m_A$ expansion,  as
\begin{equation}
\label{eq:kinetic_energy}
T_A\approx \frac{E_\nu (E_\nu-\Delta\varepsilon_{mn})(1-\cos\theta)+\Delta\varepsilon_{mn}^2/2}{m_A} \, .
\end{equation}

The differential coherent elastic neutrino-nucleus scattering (CE$\nu$NS) cross section in the BN formalism has been written as~\cite{Bednyakov:2018mjd}
\begin{equation}
\label{eq:sigma-coh1}
\begin{aligned}
\frac{d\sigma_\text{coh}}{dT_A}   = \frac{4G_F^2 m_A}{\pi}\left(1-a\right)\Bigg|\sum_{f=n,p}\sqrt{g_\text{coh}^f}F_f\Bigg(\!A^f_+\!\Big[g_L^f-g_R^fab(1-b)\Big]\!+\!A^f_-\!g_R^f\!\Big[1-ab(1-b)\!\Big]\!\Bigg)\Bigg|^2 \, ,
\end{aligned}
\end{equation}
while  the corresponding differential incoherent neutrino-nucleus scattering (I$\nu$NS) cross section reads~\cite{Bednyakov:2018mjd}
\begin{equation}
\label{eq:sigma-inc1}
\begin{aligned}
\frac{d\sigma_\text{inc}}{dT_A}   = & \frac{4G_F^2 m_A}{\pi}  \sum_{f=n,p}g_\text{inc}^f  \left(1-|F_f|^2\right) \\ & \times \Bigg[A^f_+\left(\left(g_{L,f}-g_{R,f}ab^2\right)^2+g_{R,f}^2ab^2(1-a)\right) + A^f_-g_{R,f}^2(1-a)\left(1-a+ab^2\right)
\Bigg] \, .
\end{aligned}
\end{equation}
In the latter two expressions the parameters $a$ and $b$ are defined as
\begin{equation}
a=\frac{q^2}{q^2_\text{min}}\simeq  \frac{T_A}{T_A^\text{max}}, \quad b^2=\frac{m^2_f}{s} \, .
\end{equation}
Here, $A^p_\pm \equiv Z_\pm$  ($A^n_\pm \equiv N_\pm$)  represents the number of protons (neutrons)  with spin  $\pm 1/2$ and  $s=(p+k)^2$ is the total energy squared in the center-of-mass frame ($p$ denotes an effective 4-momentum of the nucleon).  The correction factors $g_\text{coh}^f$ and $g_\text{inc}^f$ for coherent and incoherent processes, respectively, are of the order of unity and become more important for very low incoming neutrino energy (see the Appendix in Ref.~\cite{Bednyakov:2018mjd}). We have furthermore verified that, for \cevns the impact of $g_\text{coh}^f$  is less important while for the \inc channel the corresponding corrections due to $g_\text{inc}^f$ become more significant. Finally, it is interesting to recall that in the case of  zero momentum transfer $F_f(q=0)=1$ holds, and therefore, the incoherent cross section vanishes due to its dependence on the quantity  $1-F_f(q)$. 

\section{Results and discussion}
\label{sec:results}

The performed nuclear structure calculations refer to the nuclear matrix elements for $g.s. \to g.s.$ transitions (elastic channel) and for transition matrix elements in the case of inelastic scattering, for both neutrino-nucleus and WIMP-nucleus processes. 
For the elastic and inelastic WIMP-nucleus event rates given in Eqs. (\ref{eqn.12}) and (\ref{eqn.19}), the nucleonic  current part has been separated from the nuclear 
part.
We note that, $D_i$ and $E_i$ depend  on the nuclear structure
parameters as well as on the kinematics, assumptions regarding the WIMP velocity, 
and modulation effect, while $X(i)$ depend on the spin structure  functions  
and the nuclear form factors. Thus, the nuclear structure calculations are needed  for  the evaluation of $X(i)$.

\subsection{Results for WIMP-nucleus scattering}

\begin{table}[t]
\caption{Calculated static spin matrix elements $\Omega_0$ and $\Omega_1$ for elastic and inelastic channels.}
\begin{tabular}{l|ll|ll}
\hline
 Nucleus    & \multicolumn{2}{c}{Elastic} & \multicolumn{2}{c}{Inelastic} \\
\hline
          & $\Omega_0$   & $\Omega_1$   & $\Omega_0$    & $\Omega_1$    \\
$^{127}$I  & -0.494       & -0.505       & -0.276        & 0.019         \\
$^{133}$Cs & -0.878       & -0.660       & 0.020         & 0.041         \\
$^{133}$Xe & -0.835       & 0.636        & -0.031        & 0.013   \\
\hline     
\end{tabular}
\label{tab:static_ME}
\end{table}

\begin{figure*}[t]
\includegraphics[width= \textwidth]{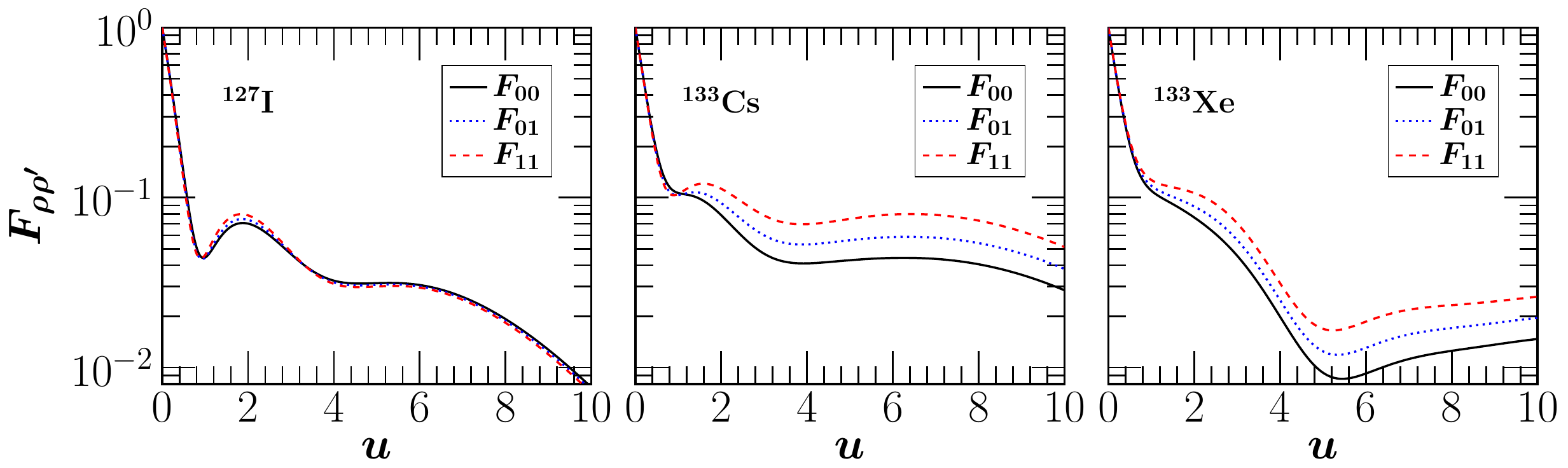}
\includegraphics[width= \textwidth]{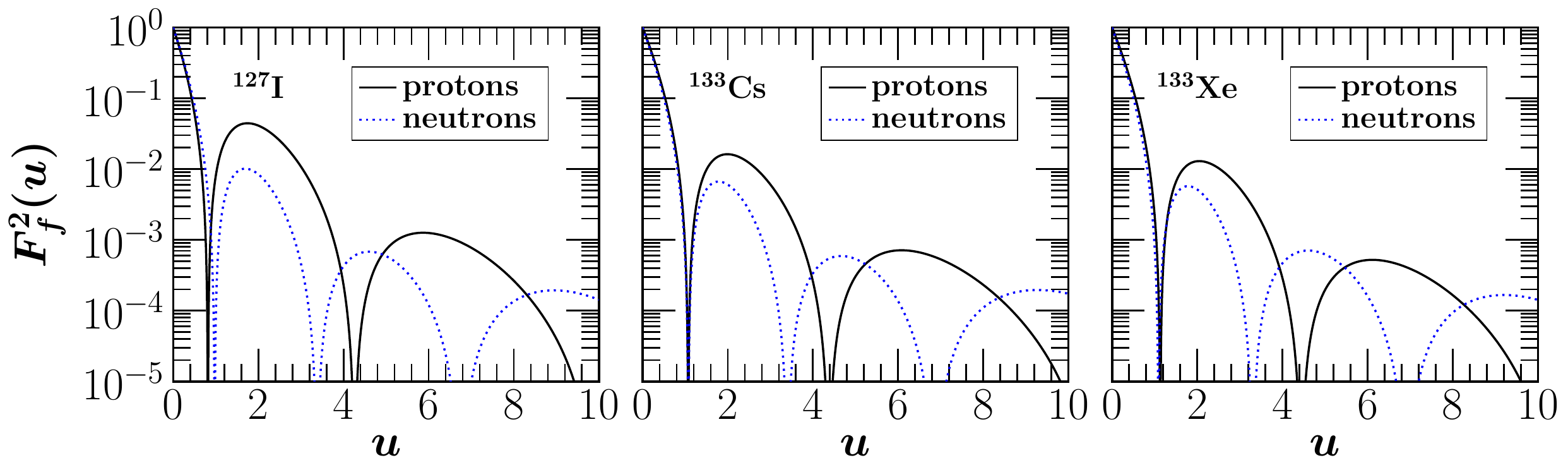}
\caption{Normalized spin structure functions (upper panels) and squared proton
and neutron nuclear form factors (lower panels) for the ground state of the $^{127}$I, $^{133}$Cs, and $^{133}$Xe isotopes.}
\label{el_ssf}
\end{figure*}

\begin{figure}[ht]
\includegraphics[width= \textwidth]{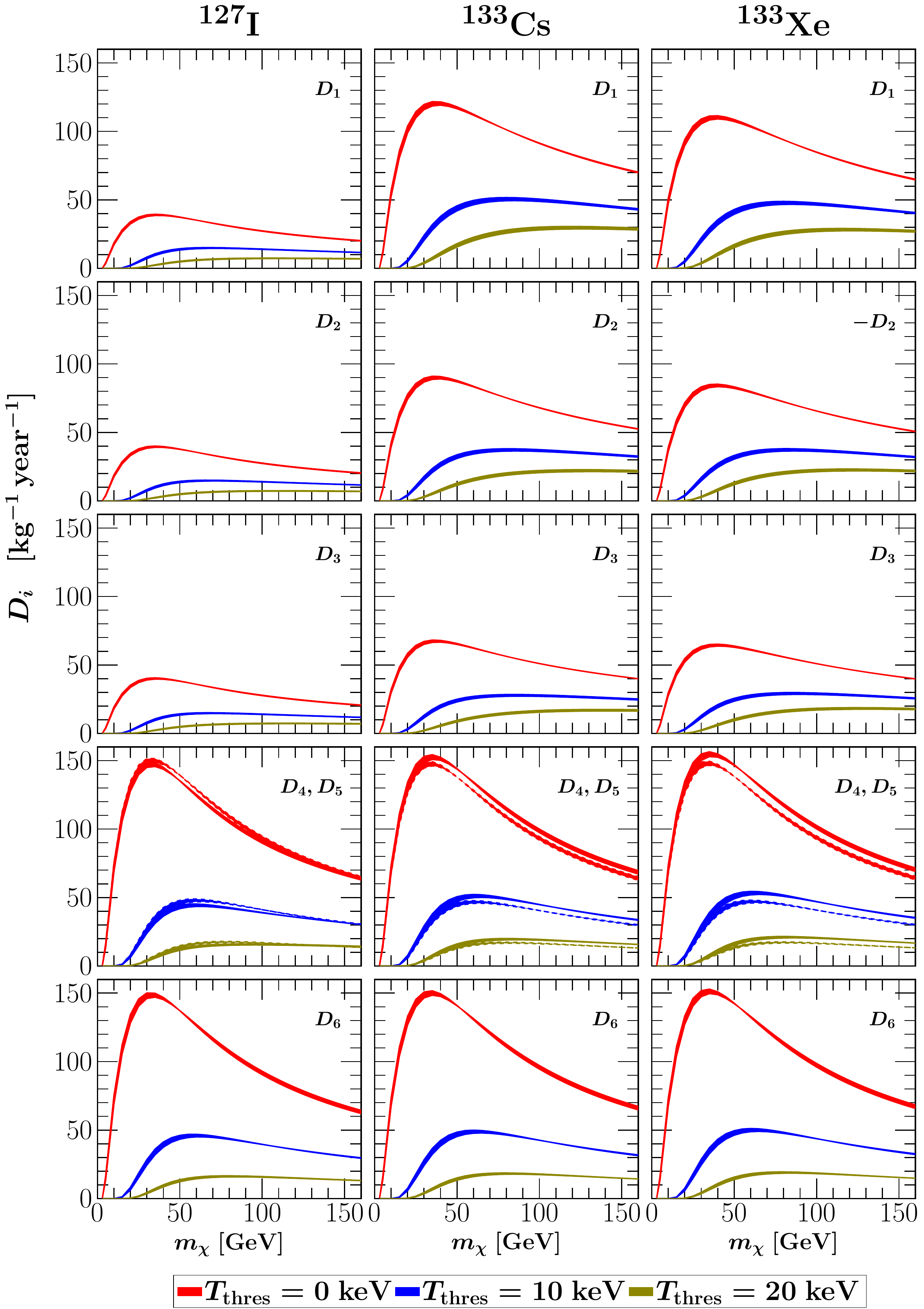}
\caption{Nuclear structure coefficients $D_i$, $i=1,\ldots, 6$,  as functions of the WIMP mass $m_\chi$ for the $^{127}$I, $^{133}$Cs, and $^{133}$Xe isotopes. The results are evaluated for three values of the detector threshold, $T_\mathrm{thres}=0,10$, and $20$ keV, while the thickness of the curves represents the annual modulation effect. Note that for the case of $^{133}$Xe the coefficient $D_2$ is negative, while the coefficients $D_5$ are shown by thick dashed curves.}
\label{dn_el}
\end{figure}

In Ref.~\cite{Papoulias:2019lfi} we have shown that the DSM reproduces better the experimental data from the COHERENT experiment (SNS neutrinos) compared to the widely used approximated methods like the Helm, symmetrized Fermi, and 
Klein-Nystrand form factors. Similarly, in Ref.~\cite{Papoulias:2018uzy}, focusing on  astrophysical neutrinos in direct detection dark-matter experiments, we showed that for a large momentum (mainly for diffuse supernova neutrino background; DSNB and atmospheric neutrinos) the difference  in the event rates for DSM vs Helm can be of one order of magnitude. This work is an extension of Refs.~\cite{Papoulias:2019lfi,Papoulias:2018uzy}, since here the inelastic channels are also considered. Compared to other relevant nuclear structure methods the DSM presents the advantage that it is applicable for odd-$A$ deformed nuclear isotopes. Finally, nuclear uncertainties in such experiments are usually taken to be 10\% in a statistical analysis, however, this issue is beyond the scope of our work.

In our calculations, we first evaluate the static spin matrix elements
$\Omega_0$ and $\Omega_1$ in Eq. (\ref{eqn.6}) for both elastic and inelastic scattering (see Table~\ref{tab:static_ME}). For the ground state of $^{127}$I these values are
-0.494 and -0.505, while for $^{133}$Cs they are -0.878 and
-0.660, and similarly, for $^{133}$Xe the corresponding quantities are -0.835 and 
0.636. Compared with other works, in Ref.~\cite{Toivanen:2009zza} the authors,  using effective $g$-factors, evaluated the elastic and inelastic event rates for $^{127}$I and found that for elastic scattering, the values of $\Omega_0^2$ and $\Omega_1^2$ 
are 0.350 and 0.226, i.e., of the same order of magnitude as in our case.
They also found that $\Omega_0$ and $\Omega_1$ have the same sign.
The DSM wave functions given by Eq. (\ref{phijm}) are used to calculate
 the normalized spin structure 
functions $F_{\rho \rho^\prime}(u)$ and the squared nuclear form
factors $|F_{Z,N}(u)|^2$ for the chosen set of nuclei. The results are shown in Fig.~\ref{el_ssf}
as a function of $u$. 
As can be seen, for the case of $^{127}$I, the structure functions  $F_{00}$, $F_{01}$, and $F_{11}$ show little variation
up to about $u=10$. 
However, for $^{133}$Cs the spin structure functions show variations from
each other above
$u=1$ (a similar trend was found in  Ref.~\cite{Toivanen:2009zza}).
Also, the DSM results for $^{133}$Xe show a similar trend.
The proton and neutron nuclear form factors for  
$^{127}$I have the same values up to about 
$u=0.8$, while beyond this value the minima of the neutron form factors are shifted towards smaller $u$ in comparison to those of protons. The form factors for $^{133}$Cs  and $^{133}$Xe
show a similar trend as in $^{127}$I. 

In Fig.~\ref{dn_el}, the nuclear-structure-dependent coefficients $D_i$, $i=1,\ldots,6$, given in Eqs. (\ref{eqn.9a})
are plotted for  $^{127}$I,  $^{133}$Cs, and $^{133}$Xe,
respectively, as functions of the WIMP mass $m_\chi$ for different values of the detector
threshold. 
The peaks of these nuclear structure coefficients occur at around $m_\chi\sim$
30 GeV in the case of zero threshold energy and they shift towards higher values of 
$m_\chi$ upon going to higher threshold energies. In this work, we studied also the crucial annual modulation effect~\cite{Bernabei:2013xsa}, which is expected to provide   strong evidence regarding the observation of DM since the background does not exhibit such modulation. In Fig.~\ref{dn_el} this effect is represented by the thickness of the graphs and it is pronounced near the peaks. For $^{127}$I, our present results have been obtained assuming threshold energies $T_\text{thres}=$
0, 10, and 20~keV, in good agreement with Ref.~\cite{Toivanen:2009zza}. It can be seen that the annual modulation effect is much smaller for this
detector nucleus, while a similar trend is seen for the other two nuclei.
On the other hand, in the case of $^{133}$Cs the values of $D_i$ are larger (by a factor of more than 2) compared to those for $^{127}$I, while in the case of $^{133}$Xe,  $D_i$ have values similar to those for $^{133}$Cs but $D_2$ is negative. Finally, the contributions to the rate driven by the coefficients $D_4$, $D_5$, and $D_6$, i.e.,  due to proton, neutron, and overlap of proton and neutron form factors, respectively, are similar for all nuclei.

\begin{figure}
\includegraphics[width=\textwidth]{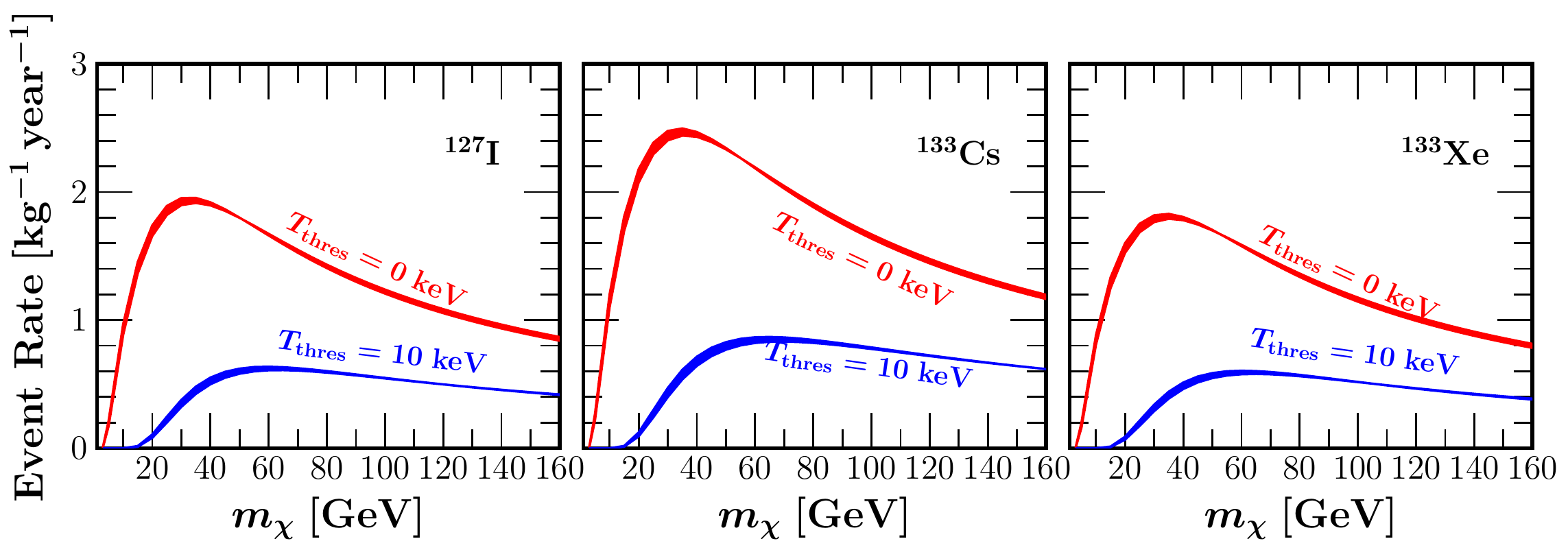}
\caption{Event rates due to elastic WIMP-nucleus scattering as a function of the
        DM mass $m_\chi$ for $^{127}$I, $^{133}$Cs, and $^{133}$Xe at detector threshold
        $T_\mathrm{thres}=0$ and $10$~keV.
        The curve thickness represents the annual modulation effect.
}
\label{rate_new}
\end{figure}

The detection event rates for the studied nuclei have been calculated at a particular
WIMP mass by reading out the corresponding values of $D_i$ in  Fig.~\ref{dn_el} for each nucleus and by using Eq. (\ref{eqn.9}) assuming the nucleonic current  parameters $f^0_A=3.55 \times 10^{-2}$,
$f^1_A=5.31 \times 10^{-2}$, $f^0_S=8.02 \times 10^{-6}$, and $f^1_S=-0.15\times f^0_S$.  Figure~\ref{rate_new} illustrates the present results as a function of the WIMP mass  $m_\chi$ for $^{127}$I,  $^{133}$Cs,
and $^{133}$Xe. The following conclusions can be drawn: for $T_\mathrm{thres}$=0 keV, the highest event rate occurs
for $^{127}$I at $m_\chi \sim 35$ GeV, while for
$^{133}$Cs and $^{133}$Xe, the highest event rates occur at $m_\chi$=40 GeV. 
The event rate decreases 
at a higher detector threshold energy but the peak shifts to  higher
values of $m_\chi$.

\begin{figure*}
\includegraphics[width= \textwidth]{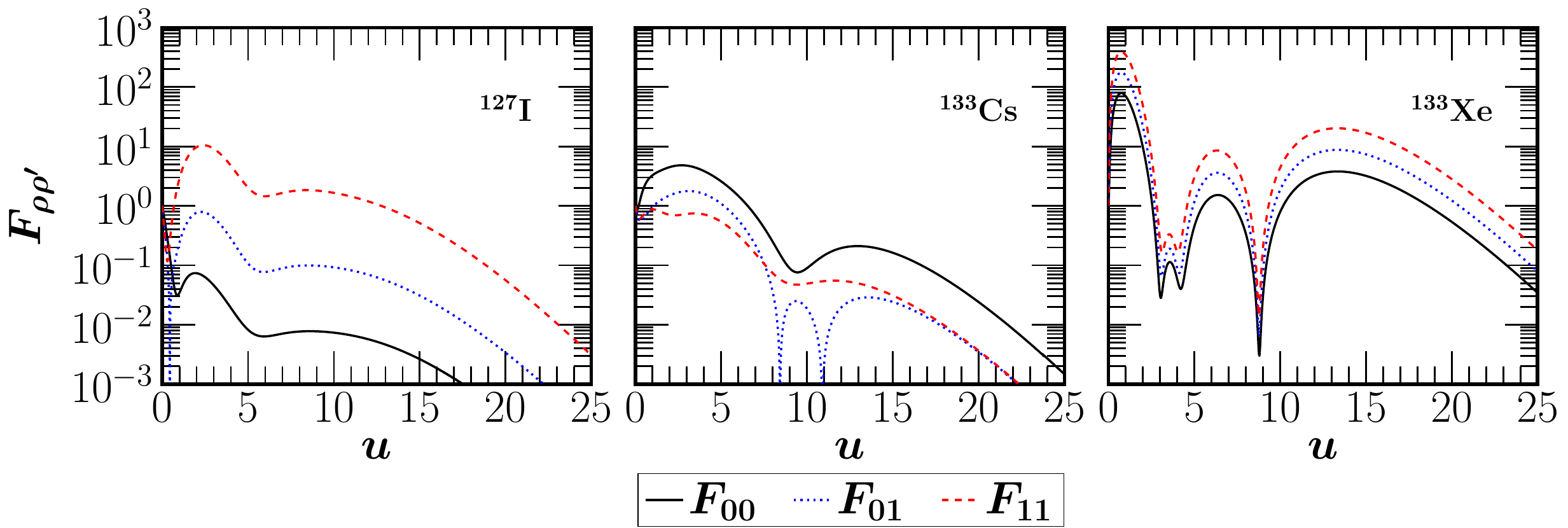}
\caption{Spin structure function   in the inelastic 
channel for $^{127}$I, $^{133}$Cs, and $^{133}$Xe.}
\label{in_ssf}
\end{figure*}

\begin{figure}
\includegraphics[width= \textwidth]{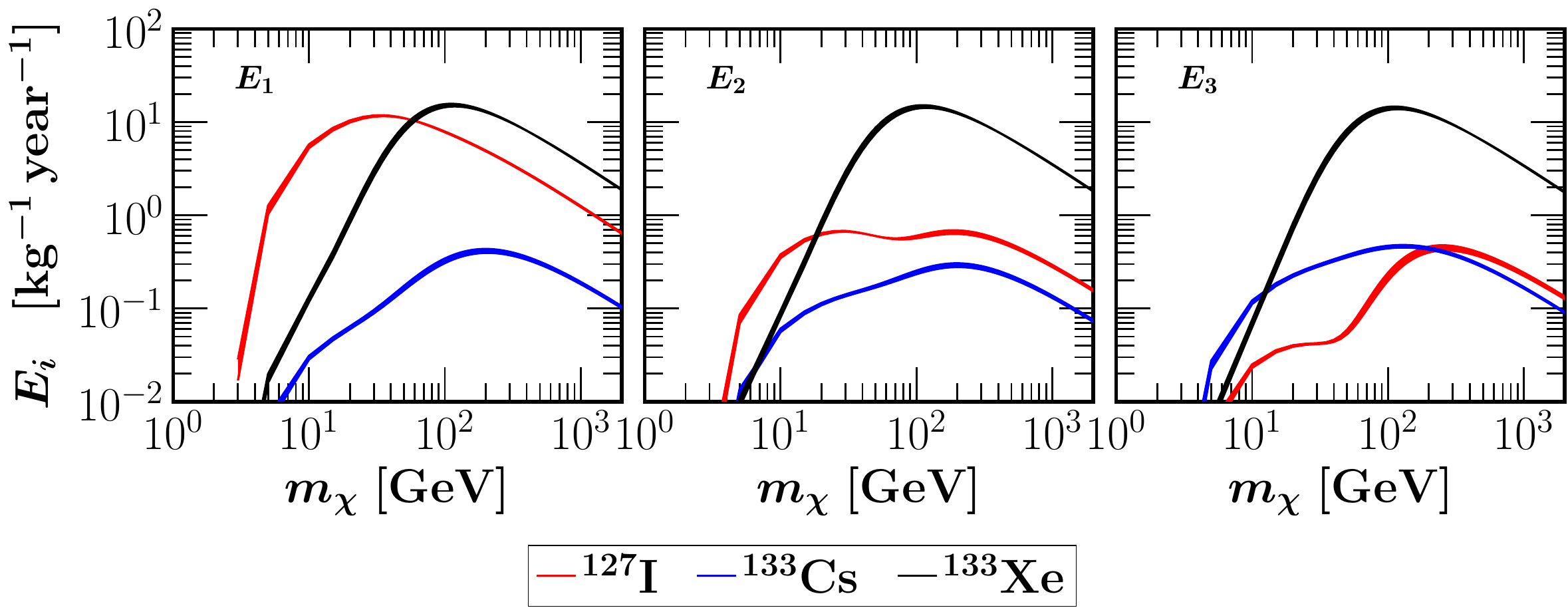}

\caption{Nuclear structure coefficients $E_i$, $i=1,2,3$, in the inelastic channel
       of $^{127}$I, $^{133}$Cs, and $^{133}$Xe.
The thickness of the curves represents the annual modulation effect. For the case of $^{133}$Xe the coefficient $E_2$ is negative.
}
\label{inel_dn}
\end{figure}

We now turn our attention to the inelastic channels of WIMP-nucleus scattering. For $^{127}$I, the first excited state $7/2^+$ appears at 57.6 keV above the ground
state. It is thus interesting to note that exotic WIMPS can potentially lead to large nucleon spin-induced cross sections,
which in turn can lead to a relatively high probability for inelastic
WIMP-$^{127}$I scattering. The latter emphasizes the significance of our present results for inelastic scattering and motivates further  study. The static spin matrix elements $\Omega_0$ and
$\Omega_1$ have values of -0.276 and 0.019, respectively. Note that unlike  the elastic
case, $\Omega_0$ and $\Omega_1$ have opposite signs.  Again, the magnitude of
$\Omega_0$ is smaller by a factor of 2 compared to the elastic case, whereas
the value of $\Omega_1$ is appreciably smaller (by  a factor of 50). 
The spin structure functions $F_{\rho \rho^\prime}$ relevant to inelastic scattering
are plotted as functions of the dimensionless momentum transfer $u$ in 
Fig.~\ref{in_ssf}. It becomes evident from the plots that the values of $F_{00}$, $F_{01}$, and $F_{11}$ are quite
different from each other. Then, the nuclear coefficients $E_i$ are plotted in Fig.~\ref{inel_dn}. At this point it is noteworthy that $E_i$ do not depend on the detector threshold energy. As in the case of elastic scattering, the corresponding  event rates due to inelastic transitions can be obtained by reading out the values
of $E_i$ from the figure and using the nucleonic current parameters and Eq. (\ref{eqn.19}).
Note that
the calculated nuclear structure coefficients are almost of the same order of 
magnitude  compared to the results given by Ref.~\cite{Toivanen:2009zza}, however, we predict a much smaller modulation effect.  Similarly, the peak values of the nuclear structure coefficients 
occur at around $m_\chi$=70--80 GeV, while in Ref.~\cite{Toivanen:2009zza} it was found that the peaks occur at
around $m_\chi\sim$200 GeV.

For $^{133}$Cs, the first excited state, having $J=5/2^+$, is at energy 81~keV. Because
of the closeness of this state to the ground state, for this nucleus also
one may expect a large nucleon spin-induced cross section, leading to a 
non-negligible probability for inelastic scattering to occur. 
The calculated static spin matrix elements $\Omega_0$ and
$\Omega_1$ have values 0.020 and 0.041, respectively. These values are almost 
40 times smaller than in the elastic case.
The spin structure functions in the inelastic channel $7/2^+\rightarrow 5/2^+$
are plotted as functions of  $u$ in Fig.~\ref{in_ssf}. The graphs for $F_{00}$, $F_{01}$ and $F_{11}$ are quite
different from each other as in the case of $^{127}$I.
As shown in  Fig.~\ref{inel_dn}, the corresponding nuclear coefficients $E_i$ for $^{133}$Cs are in general smaller compared to those of $^{127}$I. Finally, for $^{133}$Xe, the lowest excited state is $J= 1/2^+$ and appears at 262~keV. The calculated static
spin matrix elements $\Omega_0$ and $\Omega_1$ for the transition 
$3/2^+\rightarrow 1/2^+$ are $-0.031$ and $0.013$, i.e.,  more
than 30 times smaller than the elastic case in this nucleus. The
spin structure functions for this inelastic channel are plotted in Fig.
~\ref{in_ssf}, where one sees that the values of $F_{00}$, $F_{01}$, and $F_{11}$ are
quite different and show peaks. Finally, the nuclear coefficients $E_i$ are plotted 
in Fig.~\ref{inel_dn}, where it is shown that $E_1$, $E_2$, and $E_3$ are larger
than those of $^{133}$Cs, while the modulation effect is similar.

\subsection{Results for neutrino-nucleus scattering}

There are many interesting applications of low-energy neutrino scattering with nuclei that are of key importance to direct detection DM searches as shown in Ref.~\cite{Papoulias:2018uzy}.  Apart from the fact that astrophysical neutrinos pose a significant background to DM searches, neutrino-nucleus scattering experiments are also relevant and complementary to the latter for a number of reasons. Among these, we mention that  neutrino-nucleus experiments have opened a novel avenue to probe physics beyond the SM with a phenomenological impact on DM (see Ref.~\cite{Papoulias:2019xaw} and references therein). Moreover, it has recently  been shown that accelerator-produced DM can be detected in \cevns experiments~\cite{Akimov:2019xdj}. Finally, \cevns experiments can be used to probe the nuclear structure as well as to characterize the nuclear responses of the same target material employed in direct detection DM experiments.

In this section, our calculations refer to laboratory $\nu$-sources (reactor and pion decay at rest; $\pi$-DAR) and astrophysical (solar) neutrino sources having an evident connection to our previous results regarding WIMP-nucleus scattering~\cite{Papoulias:2018uzy}. We focus on popular nuclear targets employed in current and future neutrino-nucleus and DM scattering experiments. To this end, using the DSM we calculate the differential and integrated event rates due to \cevns and \inc assuming SM interactions only, as
\begin{equation}
\frac{d R_x}{d T_A} = \mathcal{K} \int_{E_\nu^\text{min}}^{E_\nu^\text{max}} \frac{d \sigma_x}{d T_A} (E_
\nu, T_A) \lambda_{\nu} (E_\nu)\, d E_\nu, \qquad x=\text{coh}, \text{inc} \, ,
\end{equation}
where $\lambda_\nu (E_\nu)$ represents the relevant neutrino energy distribution function characterizing the neutrino source.  The normalization factor $\mathcal{K}= t_\mathrm{run} \Phi_\nu N_\mathrm{targ} $ depends on the exposure time, the neutrino flux, and the number of target nuclei.  Our goal is to compare the nuclear responses between the various chosen isotopes for \cevns and \inc processes. To achieve a more direct comparison, we relax ourselves from being experiment specific, but rather in our calculations we assume the same detector mass and exposure time for all nuclear targets. Specifically, for reactor- and $\pi$-DAR-based neutrino experiments we consider 1~kg of detector mass and 1~year of data-taking, while for the case of direct DM detection experiments we assume  1~ton of detector mass and 1~year of running time.

\begin{figure}[t]
\includegraphics[width = 0.9\textwidth]{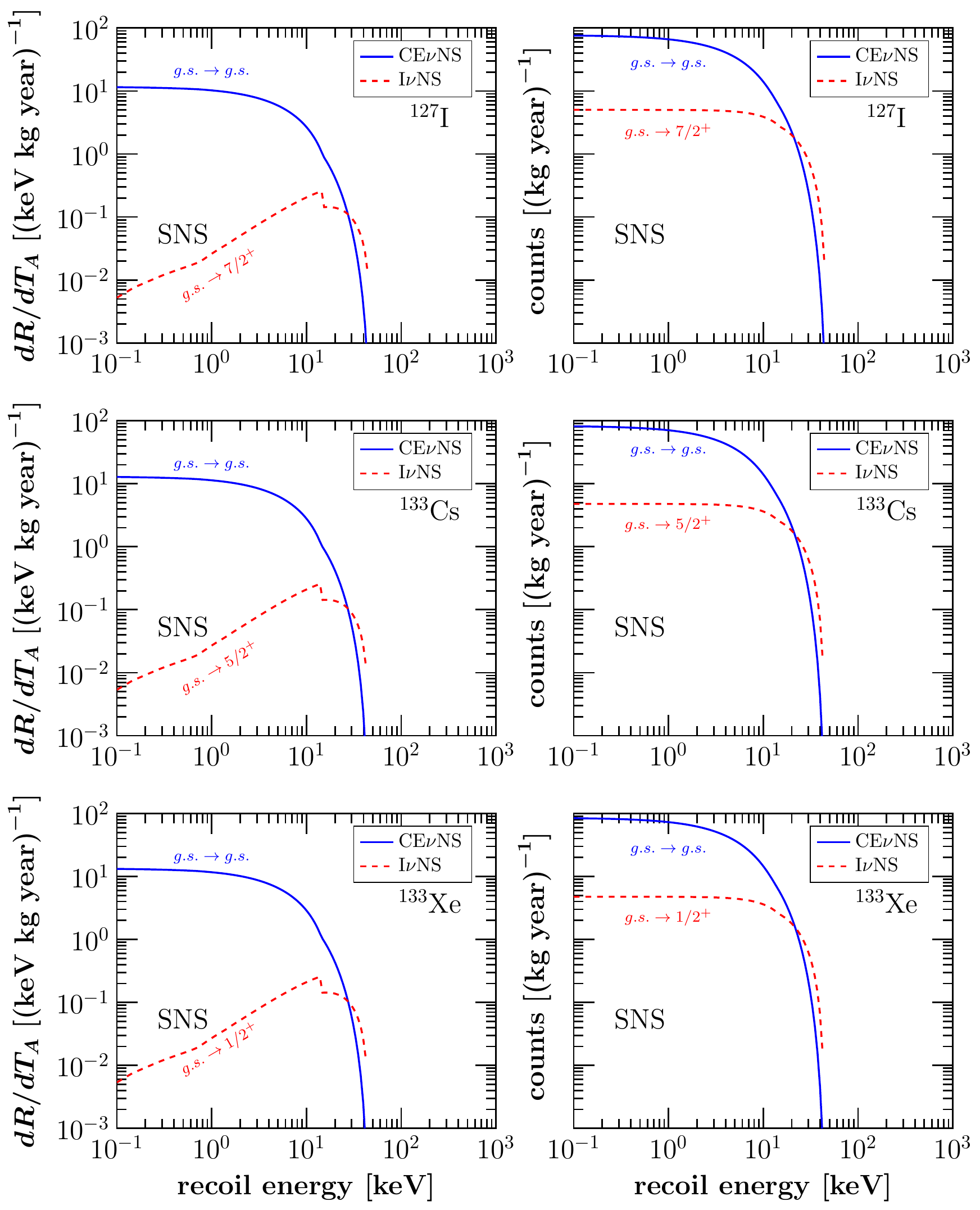}
\caption{Differential (left) and integrated (right) event rates as a function of the nuclear recoil energy for  $^{127}$I, $^{133}$Cs, and $^{133}$Xe. The results are presented for \cevns and  \inc  processes with $\pi$-DAR neutrinos.}
\label{fig:rates-SNS1}
\end{figure}

\begin{figure}[t]
\includegraphics[width = 0.9\textwidth]{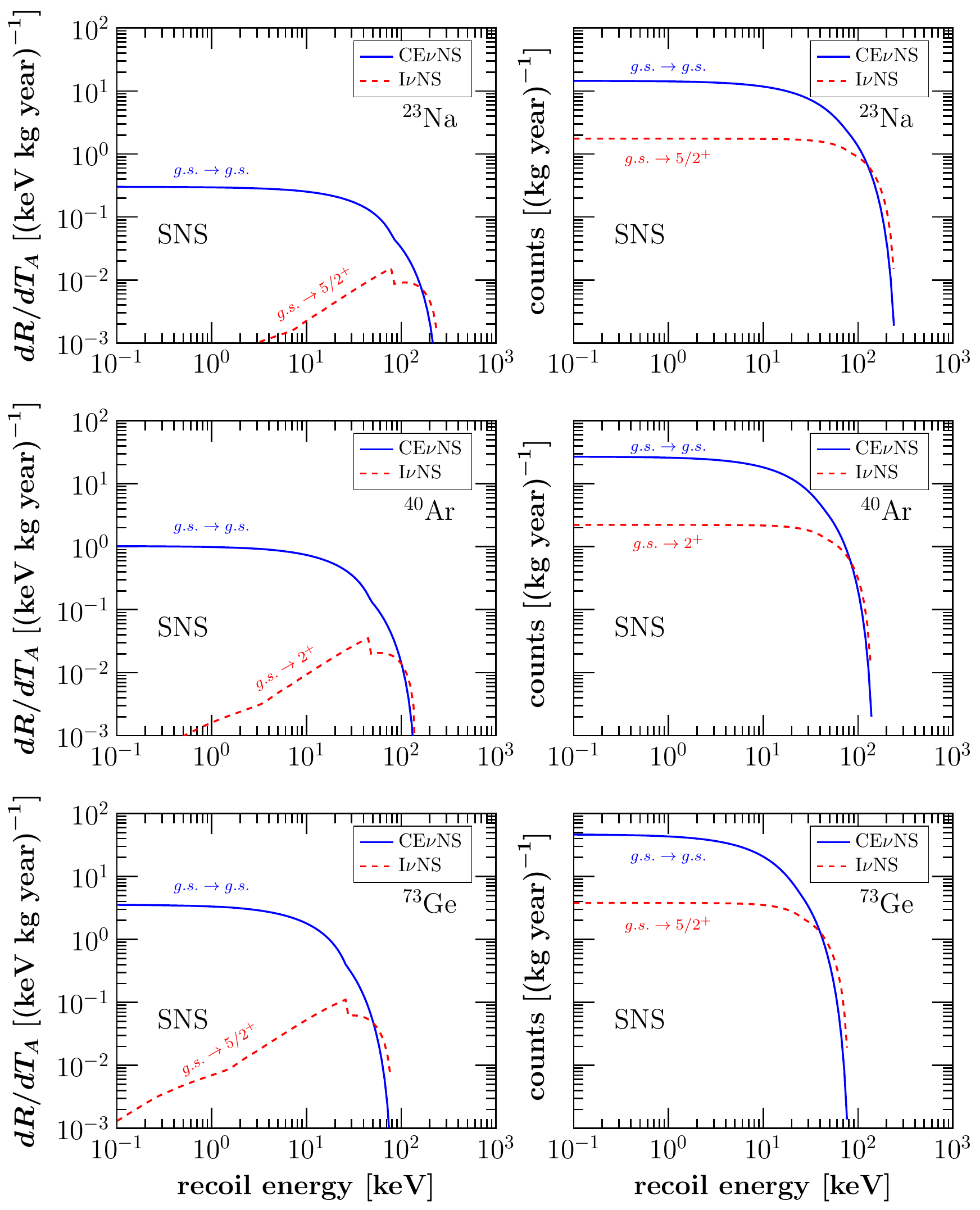}
\caption{Same as Fig.~\ref{fig:rates-SNS1}, but for $^{23}$Na, $^{40}$Ar, and $^{73}$Ge.}
\label{fig:rates-SNS2}
\end{figure}

\begin{figure}[ht]
\includegraphics[width = 0.9\textwidth]{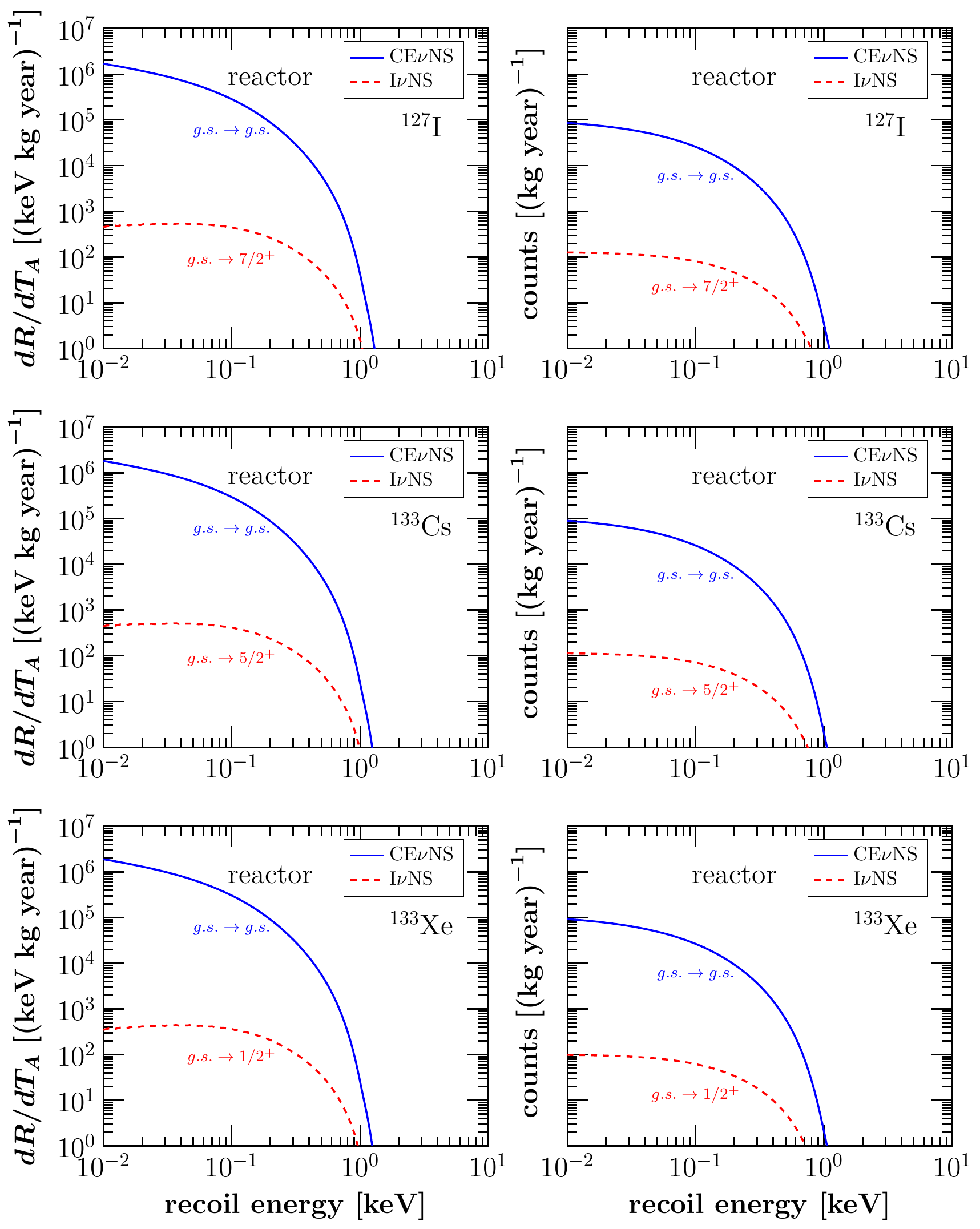}
\caption{Differential (left) and integrated (right) event rates as a function of the nuclear recoil energy for $^{127}$I, $^{133}$Cs, and $^{133}$Xe. The results are presented for \cevns and \inc processes with reactor neutrinos.}
\label{fig:rates-reactor1}
\end{figure}

\begin{figure}[t]
\includegraphics[width = 0.9\textwidth]{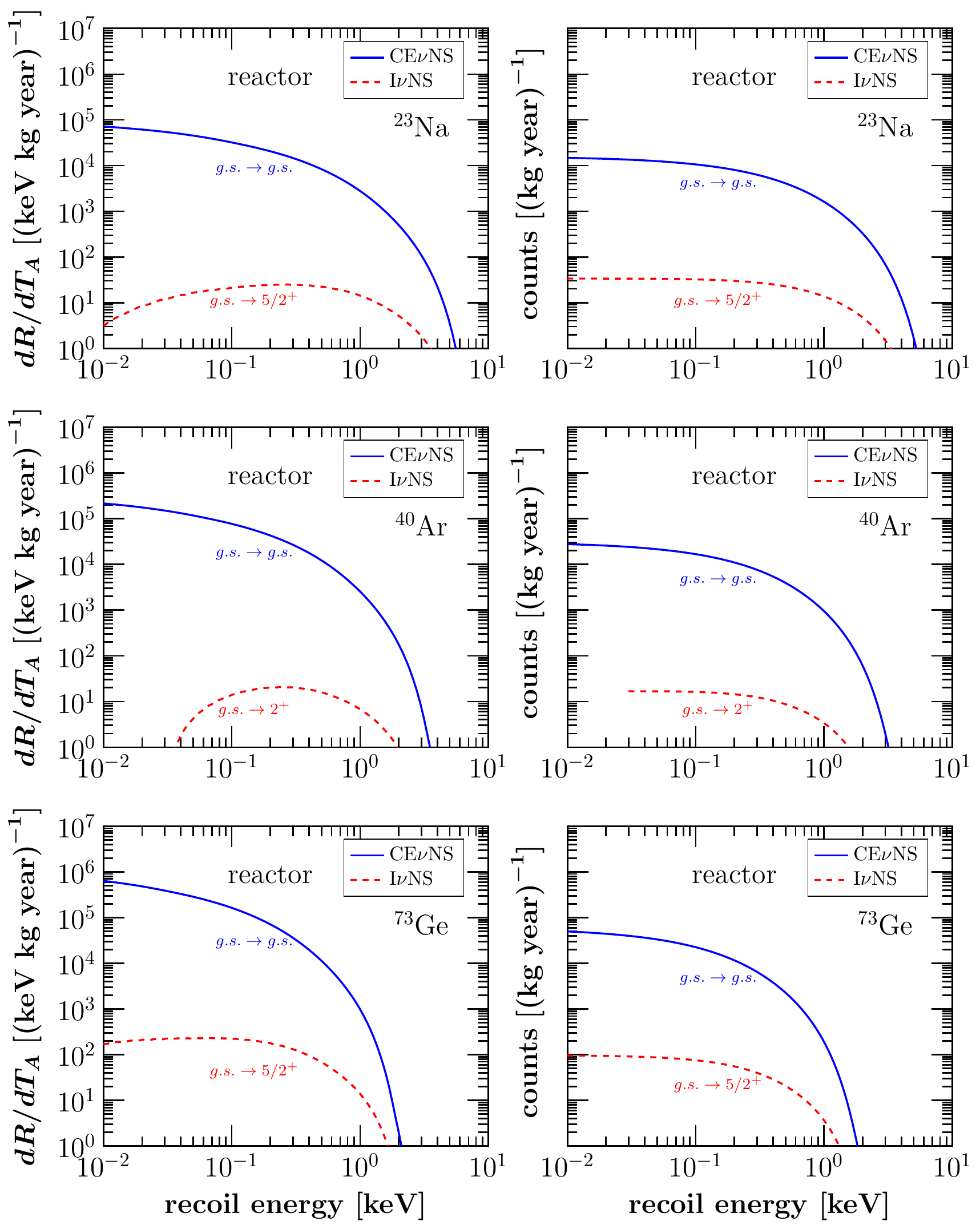}
\caption{Same as Fig.~\ref{fig:rates-reactor1}, but for $^{23}$Na, $^{40}$Ar, and $^{73}$Ge.}
\label{fig:rates-reactor2}
\end{figure}

\begin{figure}[ht]
\includegraphics[width = 0.9\textwidth]{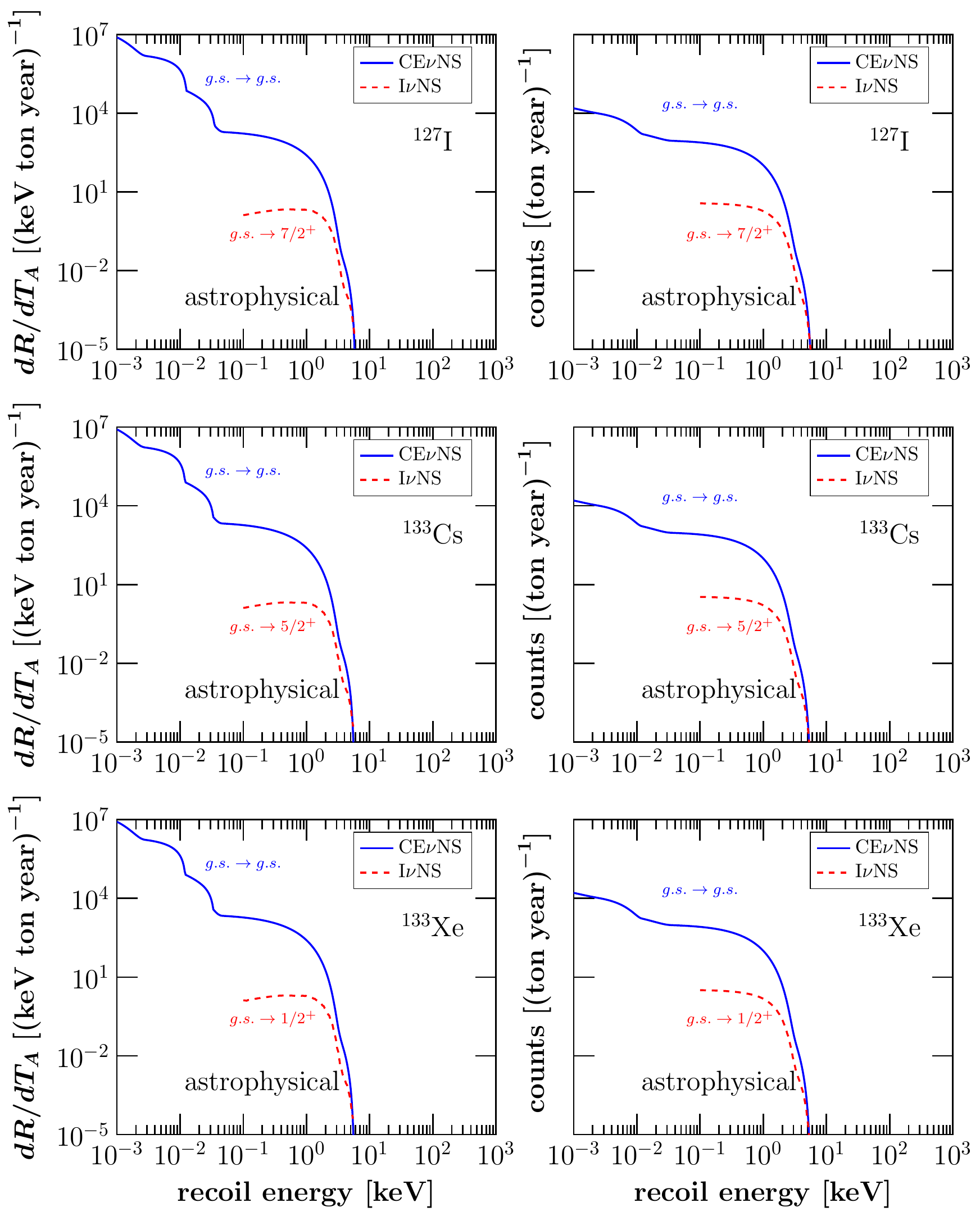}
\caption{Differential (left) and integrated (right) event rates as a function of the nuclear recoil energy for $^{127}$I, $^{133}$Cs, and $^{133}$Xe.   The results are presented for \cevns and \inc processes with solar neutrinos.}
\label{fig:rates-astro1}
\end{figure}

\begin{figure}[ht]
\includegraphics[width = 0.9\textwidth]{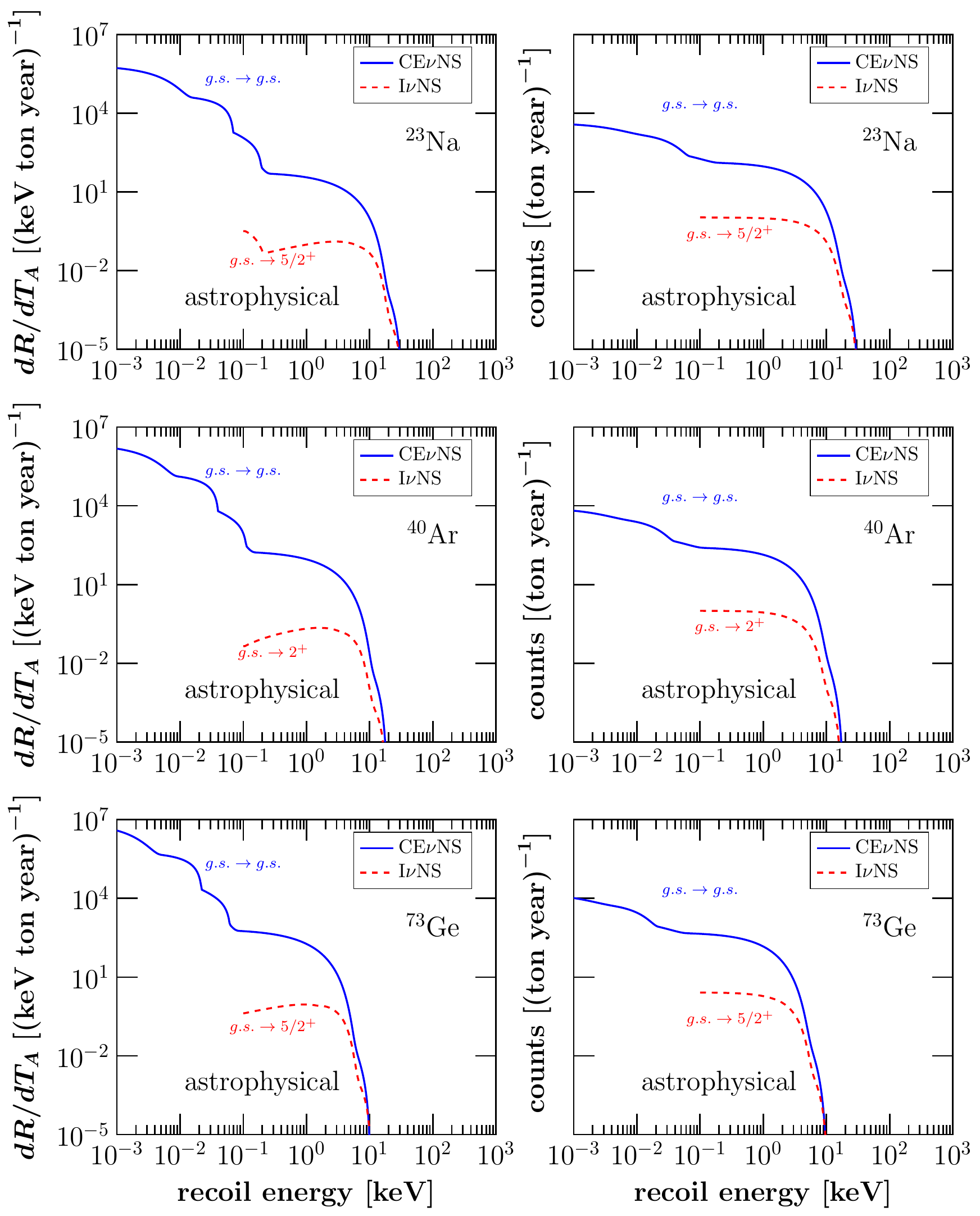}
\caption{Same as Fig.~\ref{fig:rates-astro1}, but for  $^{23}$Na, $^{40}$Ar, and $^{73}$Ge.}
\label{fig:rates-astro2}
\end{figure}

We begin our calculational procedure by considering the \cevns and \inc channels in the case of neutrino experiments utilizing $\pi$-DAR neutrinos, like the COHERENT at the SNS (the ESS~\cite{Baxter:2019mcx} is another promising possibility). It has recently been shown that a better agreement between the SM expectation and the COHERENT data is possible through DSM calculations~\cite{Papoulias:2019lfi}, while reasonable constraints on nuclear physics parameters have been placed from the COHERENT data~\cite{Cadeddu:2017etk,Miranda:2020tif}. The relevant nuclear isotopes are  $^{133}$Cs and $^{127}$I, i.e., the detector materials used during the first run of COHERENT with a CsI detector that led to the first observation of CE$\nu$NS~\cite{Akimov:2017ade,Akimov:2018vzs}. By assuming a typical SNS neutrino flux of $\Phi_\nu \sim 10^7$~$\nu \,  \mathrm{cm^{-2} s^{-1}}$ and a threshold of 100~eV we calculate the differential event rates and the number of events above threshold.  The corresponding results are presented in Fig.~\ref{fig:rates-SNS1} for $^{127}$I, $^{133}$Cs, and $^{133}$Xe.  In this figure it can be seen that the \cevns channel dominates the expected signal by at least 1~order of magnitude. However, the \inc channel can be detectable in the high-energy tail of the recoil spectrum, i.e., a possible excess in the spectral shape distribution for high recoil energies is due to potential \inc contributions. We also consider the $^{40}$Ar nucleus, which has been employed by the CENNS-10 liquid argon (LAr) detector subsystem~\cite{Akimov:2019rhz}, and recently detected \cevns with more than $3\sigma$ significance~\cite{Akimov:2020pdx}, leading to improved constraints on various parameters with regard to neutrino physics within and beyond the SM~\cite{Miranda:2020tif}. Finally, our calculations also include the $^{73}$Ge and $^{23}$Na nuclei, which are the target material of the next-generation detectors that will be deployed in the \emph{Neutrino Alley} at the SNS~\cite{Akimov:2018ghi}. The respective results for $^{23}$Na, $^{40}$Ar, and $^{73}$Ge are depicted in Fig.~\ref{fig:rates-SNS2}, leading to the same general conclusions as for the heavier isotopes. In all cases,  kinks appear at maximum recoil energies (different for each nuclear isotope) due to prompt-neutrino scattering.

We now turn our discussion to promising reactor-based experiments looking for neutrino-nucleus events. Recently, there has been a very active experimental effort with new experiments aiming to measure CE$\nu$NS~\cite{Akimov:2019wtg}. Novel developments and instrumentation based on cutting-edge technologies are currently pursued in order to reduce the detection threshold in the sub-keV region. The most interesting nuclear isotope  is $^{73}$Ge, which constitutes the main target material of the CONUS~\cite{Hakenmuller:2019ecb}, CONNIE~\cite{Aguilar-Arevalo:2019jlr}, MINER~\cite{Agnolet:2016zir}, NU-CLEUS~\cite{Angloher:2019flc}, Ricochet~\cite{Billard:2016giu}, vGEN~\cite{Belov:2015ufh}, and TEXONO~\cite{Wong:2005vg} experiments.  It is noteworthy that the $^{133}$Xe isotope is another interesting target, being the target nucleus of the RED100~\cite{Akimov:2019ogx} experiment. By assuming a reactor antineutrino flux of $\Phi_\nu \sim 10^{13}$~$\nu \,  \mathrm{cm^{-2} s^{-1}}$ we illustrate the corresponding differential and integrated event rates as a function of the nuclear recoil energy in Figs.~\ref{fig:rates-reactor1} and~\ref{fig:rates-reactor2}. Contrary  to the SNS case discussed previously, it becomes evident that the \cevns rate dominates over the \inc channel by 2--3 orders of magnitude. As expected, the \inc channel is not relevant due to the very low momentum transfer occurring when low-energy reactor antineutrinos scatter off nuclei, i.e., the $1-F_f(|\mathbf{q}|)$ factor becomes tiny, leading to an appreciable suppression of the incoherent cross section.

Finally, we study \cevns and \inc processes due to nuclear interactions of neutrinos at ton-scale direct DM detectors. The latter constitute an irreducible background to WIMP-nucleus scattering events in direct detection DM searches, also known as the neutrino-floor~\cite{Billard:2013qya}. Indeed, a neutrino-induced event can mimic a potential WIMP-nucleus signal, which may subsequently lead to an erroneous interpretation of DM observation. Moreover, it has been noted recently that existing uncertainties in the SM event rate, mainly those coming from the nuclear form factor, require further attention~\cite{Papoulias:2018uzy,AristizabalSierra:2019zmy}. For this reason, in this work a comprehensive calculation that takes into account realistic nuclear structure calculations through the DSM as well as the nuclear responses from both \cevns and \inc interactions, is performed. Furthermore, since we are interested in low-threshold detectors we restrict our selves to solar neutrinos only, without including the contributions due to atmospheric and DSNB  neutrinos. The combined effect of the large energy-related uncertainties characterizing the atmospheric and DSNB neutrino spectra, in conjunction with the corresponding ones regarding the form factors at large momentum-transfer, stand out as another motivation for this assumption. The target nuclei of interest are $^{40}$Ar (DarkSide~\cite{Agnes:2018oej}, DEAP-3600~\cite{Amaudruz:2017ekt}), $^{73}$Ge (CDEX, SuperCDMS~\cite{Agnese:2016cpb}), and $^{133}$Xe (LUX~\cite{Akerib:2016vxi}, XENON1T~\cite{Aprile:2015uzo}, DARWIN~\cite{Aalbers:2016jon}, PandaX-II~\cite{Fu:2016ega}). Our DSM results are presented in Figs.~\ref{fig:rates-astro1} and~\ref{fig:rates-astro2}.  As expected, the \inc rate coming from the low-energy solar neutrino spectra is subdominant with respect to the \cevns rate. On the other hand, this may not be the case when atmospheric and DSNB neutrinos are taken into account. As mentioned previously, the expected \inc rates for high-energy neutrinos and at high momentum transfer require special attention that goes beyond the scope of this study. Such results are in progress and will be published in a future work.

\section{Summary and Conclusions}
\label{sec:conclusions}

In this study we have presented the expected neutrino-nucleus and WIMP-nucleus events at prominent rare-event detectors, calculated in the framework of the deformed shell model. The chosen nuclear isotopes, $^{23}$Na, $^{40}$Ar, $^{73}$Ge, $^{127}$I, $^{133}$Cs, and $^{133}$Xe, were carefully selected to correspond to present and future experiments looking for \cevns events and WIMP candidates at direct detection dark-matter detectors. 

The nuclear effects being probably the largest source of theoretical uncertainty are limiting the potential of the relevant experiments in placing constraints on physics beyond the SM. In this work, the addressed corrections coming from nuclear structure are adequately taken into consideration through the DSM and are essential to accurately simulate the expected event rates. In particular, on the basis of Hartree-Fock nuclear states, we have conducted a comprehensive study by also taking into account the deformation of the studied nuclear isotopes, leading to a more accurate determination of the spin nuclear structure function and the nuclear form factors for protons and neutrons. 

We have, furthermore, illustrated a comparison of the relevant magnitude between the coherent and the incoherent processes as well as their recoil energy dependence assuming various neutrino sources. Our obtained results indicate that incoherent neutrino-nucleus processes can lead to an observable enhancement of the expected signal that is well-above the energy threshold at multi-ton direct detection dark-matter detectors. Similarly a corresponding enhancement was found for the case of WIMP-nucleus scattering. We have finally emphasized that in both cases a signal coming from the de-excitation gammas in the aftermath of an incoherent processes can be erroneously misinterpreted as a possible new physics signature.

\acknowledgments
The research of DKP is co-financed by Greece and the European Union (European Social Fund- ESF) through the Operational Programme «Human Resources Development, Education and Lifelong Learning» in the context of the project ``Reinforcement of Postdoctoral Researchers - 2nd Cycle" (MIS-5033021), implemented by the State Scholarships Foundation (IKY).
RS is thankful to SERB of Department of Science and Technology (Government of
India) for financial support. The work of TSK is implemented through the Operational Program ``Human Resources Development,
Education and Lifelong Learning - cycle B''  (MIS-5047635) and is co-financed by the European Union (European Social Fund) and Greek national funds.

\bibliographystyle{utphys}
\bibliography{bibliography}

\end{document}